\documentclass[12pt]{article}
\usepackage[dvips]{graphicx}
\usepackage{dsfont}
\usepackage{amssymb}
\usepackage{amsmath}
\usepackage{epsfig}
\usepackage{cite}
\usepackage{hyperref}
\usepackage{bbold}
\usepackage{multirow}
\usepackage{mathtools}

\numberwithin{equation}{section}
\numberwithin{table}{section}

\def\beq{\begin{equation}}
\def\eeq{\end{equation}}
\def\be{\begin{equation}}
\def\ee{\end{equation}}
\def\bea{\begin{eqnarray}}
\def\eea{\end{eqnarray}}
\def\d{{\rm d}}
\DeclareRobustCommand{\SkipTocEntry}[4]{}
\RequirePackage{color}

\newcommand{\cM}{\mathcal{M}}
\newcommand{\cN}{\mathcal{N}}

\newcommand{\cV}{\mathcal{V}}

\newcommand{\dd}{\mathrm{d}}
\newcommand{\D}{\mathrm{D}}

\newcommand{\cVT}{{\mathcal{V}_{\scalebox{0.5}{T}}}}
\newcommand{\tcVT}{{\tilde{\mathcal{V}}_{\scalebox{0.5}{T}}}}
\newcommand{\cVH}{{\mathcal{V}_{\scalebox{0.5}{H}}}}
\newcommand{\tcVH}{{\tilde{\mathcal{V}}_{\scalebox{0.5}{H}}}}
\newcommand{\PT}{P_{\scalebox{0.5}{T}}}
\newcommand{\PH}{P_{\scalebox{0.5}{H}}}
\newcommand{\PHi}{P_{\scalebox{0.5}{H}\,i}}
\newcommand{\QT}{Q_{\scalebox{0.5}{T}}}
\newcommand{\QH}{Q_{\scalebox{0.5}{H}}}
\newcommand{\DT}{D_{\scalebox{0.5}{T}}}
\newcommand{\dH}{D_{\scalebox{0.5}{H}}}

\def\SU{{SU}}

\def\SO{{SO}}

\def\USp{{U\! Sp}}
\def\ee{{\mathfrak{e}}}
\def\E{{E}}
\def\Spin{{\mathrm{Spin}}}
\def\ie{{\it i.e.\ }}
\def\eg{{\it e.g.\ }}

\begin{document}

\begin{titlepage}
\begin{center}
\rightline{\small }

\begin{flushright} 
CPHT-RR023.052019 \\
IPhT-T19/019
\end{flushright}

\vskip 2cm

{\Large \bf Microstate geometries \\ at a generic point in moduli space}
\vskip 1.2cm

{ Guillaume Bossard$^{a}$\footnote{E-mail: guillaume.bossard@polytechnique.edu} and  
Severin L\"ust$^{b,a}$\footnote{E-mail: severin.luest@polytechnique.edu} }
\vskip 0.3cm
{\small\it  $^{a}$ Centre de Physique Th\'eorique, CNRS,  Institut Polytechnique de Paris,  \\ 
91128 Palaiseau Cedex, France} \\
\vskip 0.2cm
{\small\it  $^b$ Institut de Physique Th\'eorique, 
Universit\'e Paris Saclay, CEA, CNRS\\
Orme des Merisiers \\
91191 Gif-sur-Yvette Cedex, France} \\
\vskip 0.8cm

{\tt }

\end{center}

\vskip 1cm

\begin{center} {\bf Abstract }\\

\end{center}

\vspace{0.2cm}

We systematically study all supersymmetric solutions of six-dimensional  $(2,0)$ supergravity  with a null isometry.
In particular, every such solution with at least four real supersymmetries is also a supersymmetric solution of a $(1,0)$ theory preserving the same absolute amount of supersymmetry. 
This implies that no genuinely new solutions of this type can be found in this framework. The microstate geometries associated to supersymmetric black holes within Mathur's proposal are generically supersymmetric solutions of six-dimensional supergravity. A direct consequence of our result is that supersymmetric microstate geometries of single centre supersymmetric black holes should carry only one compact 3-cycle.

\noindent

\vfill


\end{titlepage}


\setcounter{tocdepth}{1}
\tableofcontents

\section{Introduction}

One of the most fascinating problems in quantum gravity is the black hole information paradox~\cite{Hawking:1976ra,Mathur:2009hf}. In string theory the Bekenstein--Hawking entropy is understood to be associated with an exponentially large number of states which can be interpreted as string excitations with D-brane boundary conditions in a weakly coupled regime~\cite{Strominger:1996sh}. This, by itself, does not yet provide a resolution of the paradox. Nevertheless,  there are indications that in string theory the physics at the horizon scale may be sufficiently modified by quantum effects to resolve the paradox~\cite{Mathur:2005zp,Bena:2007kg,Skenderis:2008qn,Balasubramanian:2008da,Mathur:2012zp,Bena:2013dka}.

Mathur proposed that the exponentially large number of accessible quantum states allows for non-negligible quantum effects, despite the fact that the Riemann tensor measured in Planck units is very small in the vicinity of the horizon of a large black hole \cite{Mathur:2009hf}. He furthermore proposes that one may be able to test this hypothesis semi-classically in some regime of superstring theory \cite{Lunin:2001jy,Mathur:2005zp}. 

For a  classical globally hyperbolic solution in supergravity, one may consider that there exists a quantum gravity state that is well approximated on a Cauchy surface $\Sigma$ by a Dirac distribution type wave functional, peaked on the pullback of this classical solution onto the Cauchy surface. When quantum effects are negligible, the evolution of such a state should be such that it is defined by the time evolution of the classical solution itself, so that in particular, a stationary solution would correspond to a stationary state. This picture cannot  be directly applied to a black hole, which is not globally hyperbolic by essence. But a slight generalisation of the same picture is to consider instead a quantum state that is a linear superposition of microstates which would themselves be such Dirac distribution type wave functionals peaked on slices of globally hyperbolic solutions. From this point of view, a black hole would be in a quantum superposition of a very large number of microstates which could individually be described semi-classically as peaked distributions on classical smooth globally hyperbolic solutions. The microstate geometries associated to a given black hole are constrained to have the same asymptotic charges as the black hole solution. 

If one could define the whole set of microstate geometries associated to a given black hole, one could in principle compute any observable of the black hole as a quantum average over some distribution of the same amplitude computed in the background of each microstate separately. For example, one would expect that the gravitational attraction of a microstate geometry does not necessarily capture a test particle for generic incoming boundary conditions \cite{Bianchi:2017sds}.
The black hole, on the other hand, captures the test particle if the impact parameter is small enough. The average of the evolution over a large enough set of microstate geometries should therefore reproduce the capture for a small enough impact parameter. Ideally, all classical observables computed in the microstate geometry should reproduce, after average, the observables computed in the black hole background to an excellent approximation.

However, it is not clear that one can construct a complete basis of microstates as such peaked distribution on microstates geometry. And even if this were possible, one may wonder if the typical solutions do not involve arbitrary small cycles such that one could not simply use them as classical backgrounds without neglecting quantum corrections. For this purpose it was suggested in  \cite{Bena:2013dka} to distinguish between microstate geometries, that are differentiable manifolds with an everywhere small Riemann tensor in Planck units, from microstate solutions, that possibly involve mild singularities which can be resolved in perturbative string theory, like orbifold singularities, and a bounded Riemann tensor, but not necessarily small in Planck units. For microstate geometries one can use supergravity as an effective theory, while for microstate solutions one needs to take string theory corrections into account. They also propose a third class, the fuzz-balls, that would be genuinely non-perturbative string theory quantum states for which there is no supergravity approximation. A large class of these geometries has been built explicitly. One can distinguish two main types of solutions. Firstly, the multi-bubble solutions that  are regular in five dimensions and usually involve several cycles \cite{Bena:2007kg,Giusto:2004id,Bena:2005va,Berglund:2005vb,Bena:2006kb,Bena:2007qc,Bianchi:2017bxl,Heidmann:2017cxt,Bena:2017fvm}, and secondly, the supertube solutions that  usually only involve one cycle and are regular in six dimensions \cite{Lunin:2012gp,Giusto:2013bda,Bena:2015bea,Bena:2016agb,Bena:2016ypk,Bena:2017xbt}. Both types can also be combined, but this has not been done explicitly for the co-dimension one generic solutions \cite{Bena:2015bea,Bena:2016agb,Bena:2016ypk,Bena:2017xbt}, which are called superstrata. One may also consider solutions that are only smooth locally, but for which the patching between open sets would involve U-duality transformations \cite{Park:2015gka,Fernandez-Melgarejo:2017dme,Hull:2004in,Hellerman:2002ax}. The original supertube solutions are microstate geometries describing small black holes that have a vanishing horizon in classical supergravity. It has been exhibited that their typical microstates solutions involve arbitrarily small cycles for which one cannot trust the supergravity approximation \cite{Martinec:1999sa,Lunin:2002iz,Lunin:2002qf,Chen:2014loa,Marolf:2016nwu}. One may wonder if the same problem occurs for three-charge black hole that carry a microscopic horizon, but one may hope that arbitrarily small cycles are only required when the horizon itself is small in string scale units.

For a given black hole, a microstate geometry is defined as a globally hyperbolic smooth solution with the same asymptotic charges as the black hole. One should in principle wonder about the uniqueness of the black hole solution. For a non-extremal asymptotically Minkowski black hole in four dimensions, the solution is uniquely determined by its total mass, angular momentum and electromagnetic charges. But for BPS black holes one must distinguish a single centre black hole with a unique horizon from bound states of black holes that carry the same electromagnetic charges and total energy \cite{Denef:2007vg}.  The situation is even more complicated for black holes in five dimensions, because there the horizon can admit different topologies \cite{Emparan:2001wn,Elvang:2007rd}, and black objects can also be surrounded by topological cycles carrying flux \cite{Kunduri:2014iga}. In four dimensions a BPS black hole has no angular momentum, so a microstate geometry associated to the black hole should have no angular momentum either, or at least one should find that the angular momentum vanishes in average on the set of microstate geometries. But there are also multi-centre black hole bound states without angular momentum, and one must distinguish their microstate geometries. Such black hole bound states have the property that they do not exist for arbitrary asymptotic values of the scalar fields.
Therefore, the microstate geometries associated to the single-centre black hole can be distinguished by the requirement that they exist everywhere in moduli space.

It was indeed exhibited in \cite{Dabholkar:2009dq} that multi-centre black hole solutions with at least two black holes with a large horizon do not exist at a generic point in moduli space. This property follows from the fact that the most general asymptotically flat supersymmetric solution of (ungauged)  $\cN\ge 4$ supergravity in four dimensions with four supersymmetries and a time-like isometry must also be a  solution of an $\cN=2$ truncation \cite{Bossard:2010mv,Meessen:2010fh}. This is very important for precise counting of black hole microstates in four dimensions, because only bound states of two 1/2 BPS black holes contribute to the helicity supertrace \cite{Dijkgraaf:1996it,Dijkgraaf:1996xw}, so that wall crossing corrections are necessarily small and can accurately be separated from the single centre black hole microstate counting \cite{Dabholkar:2012nd}.

For a supersymmetric black hole, which type of microstate geometries preserving the same  supersymmetry exist at generic values of the moduli? This is the question we shall investigate in this paper.  For this purpose we consider the effective supergravity description for type IIB string theory on $T^4$ or K3, which is a supergravity theory in six spacetime dimensions of type (2,2) or (2,0) respectively. In the maximal case, we shall only consider the maximal (2,0) truncation, disregarding the vector fields. For a four-dimensional supersymmetric black hole, the microstate geometries are asymptotically four-dimensional Minkowski space times two circles fibred over the Minkowski base. The electromagnetic charges of the four-dimensional black hole originate geometrically from the momentum along the two circles and their fibration over the base, as well as from the 3-form fluxes of the six-dimensional 3-form field strengths. The 3-form fluxes are supported by 3-cycles of the Euclidean base space of the microstate geometry, and in string theory each individual flux is constrained to belong to an even-selfdual lattice. We shall find that the solutions only exist for generic values of the asymptotic scalar fields if all the fluxes associated to the different 3-cycle are proportional to each other, and therefore rational multiples of the total flux associated to the black hole. This seems to mean that the superstratum microstate geometries found in \cite{Bena:2015bea,Bena:2016agb,Bena:2016ypk,Bena:2017xbt} indeed describe black holes, while the multi-bubble type microstate geometries would rather correspond to bound states of black holes.
Multi-bubble geometries always involve more than one cycle with linearly independent fluxes and therefore do not exist everywhere in moduli space.

Supersymmetric solutions of supergravity theories have been characterized for the first time in \cite{Tod:1983pm} following results of \cite{Gibbons:1982fy} for the case of pure $\cN = 2$, $d = 4$ supergravity by assuming the existence of a Killing spinor and constructing bosonic objects from its bilinears.
In six dimensions this method was first applied in \cite{Gutowski:2003rg} on pure $ (1,0)$ supergravity, not coupled to any matter multiplets.
This work was later extended to Fayet-Iliopoulos-gauged supergravity with vector multiplets and one tensor multiplet in \cite{Cariglia:2004kk}, to ungauged supergravity with an arbitrary number of tensor multiplets in \cite{Lam:2018jln} and finally to ungauged supergravity with vector, tensor as well as hypermultiplets in \cite{Cano:2018wnq}.
It was discovered in \cite{Bena:2011dd} that the underlying equations of these solutions exhibit a linear structure, and specific solutions have been constructed in \cite{Martelli:2004xq, Bobev:2012af, Niehoff:2012wu, Niehoff:2013kia, Giusto:2013rxa, Bena:2015bea, deLange:2015gca, Bena:2017xbt}.
For the six-dimensional $(2,0)$ theories, on the other hand, so far only maximally supersymmetric solutions have been classified \cite{Chamseddine:2003yy}, see also \cite{Louis:2016tnz}.

It is one of the aims of this paper to fill this gap and to classify supersymmetric solutions of six-dimensional $(2,0)$ supergravity coupled to an arbitrary number of tensor multiplets.
Notice, that this theory does not allow for any gaugings or massive deformations \cite{Bergshoeff:2007vb}.
In particular the former can easily be seen from the absence of any vector fields.
Therefore, this is already the most general $ (2,0)$ theory.
Supersymmetric solutions of supergravity theories admit at least one isometry. 
The associated Killing vector field $V$ can be obtained as a bilinear of the Killing spinor $\epsilon$,
\ie $V^\mu = \bar \epsilon \gamma^\mu \epsilon$.
In the case at hand this isometry can be either time-like or null, corresponding to $V \cdot V > 0$ or $V \cdot V = 0$. Black hole solutions in five or four dimensions uplift to supersymmetric solutions in six dimensions with a null isometry. Therefore, it is the second case of a null (\ie light-like) isometry that is relevant to microstate geometries, and hence this is the case we discuss in this paper. Here, we find that every supersymmetric solution with four preserved supercharges (of the type allowing for a black hole solution)  is at the same time always also a supersymmetric solution of a $(1,0)$ theory, preserving the same absolute amount of supersymmetry.
Hence, it will not be possible to find any genuinely new  solutions with the same asymptotic metric as a supersymmetric black hole.

This paper is organized as follows.
In Section~\ref{sec:20sugra} we review $ (2,0)$ supergravity in six dimensions. 
In Section~\ref{sec:susysolutions} we study its supersymmetric solutions and show that all solutions with a null isometry are equivalent to solutions of a $(1,0)$ theory.
Section~\ref{sec:microstates} discusses some implications on the construction of microstate geometries.
In Section~\ref{sec:chargequantization} we finally show that at a generic point in moduli space every solution has parallel fluxes.

\section{Six-dimensional \texorpdfstring{$(2,0)$}{(2,0)} supergravity}\label{sec:20sugra}

In this section we review the relevant properties of  $(2,0)$ six-dimensional, \ie chiral (ungauged) half-maximal supergravity, which has been constructed in \cite{Romans:1986er, Riccioni:1997np}.

The field content of the theory includes one gravity multiplet coupled to \(n\) tensor multiplets, which decompose as
\begin{equation}
\left(g_{\mu\nu}, \psi^A_\mu, B^I_{\mu\nu},  \chi^{Ar}, \cV^a{}_I\right) \,,
\end{equation}
where \(g_{\mu\nu}\) is the metric and \(\psi^A_\mu\), \(A = 1,\dots,4\), are symplectic Majorana--Weyl gravitini, transforming in the fundamental representation of the R-symmetry group \(\USp(4)\).
The \(B^I_{\mu\nu}\), \(I = 1,\dots,5+n\), are chiral tensor fields and transform as a vector under the global symmetry group \(\SO(5,n)\).
Their field strengths \(G^I = \dd B^I\) satisfy a twisted selfduality equation that we shall display shortly. 
The spin-1/2 fermions \(\chi^{Ar}\), \(r = 1, \dots, n\), are in the fundamental representation of \(\USp(4)\) as well as of \(\SO(n)\).
All fermions in the theory are chiral, we have
\begin{equation}
\gamma_7 \psi^A_\mu  = - \psi^A_\mu \,,\qquad \gamma_7 \chi^{Ar} = \chi^{Ar} \,.
\end{equation}
The tensor multiplets include $5 n$ scalar fields that parametrize the coset manifold
\begin{equation}\label{eq:scalarmanifold}
\cM = \frac{\SO(5,n)}{\SO(5) \times \SO(n)} \, , 
\end{equation}
through a coset representative in  \(\SO(5,n)\),
\begin{equation}
\cV = \left( \cV^a{}_I, \cV^r{}_I \right) \,,\qquad I = 1, \dots, 5 + n \,.
\end{equation}
The coset representative satisfies
\begin{equation}\label{eq:etaA}
\eta_{IJ} = \delta_{ab}\cV^a{}_I \cV^b{}_J -\delta_{rs}  \cV^r{}_I \cV^s{}_J \,,
\end{equation}
where $\eta_{IJ}$ is a metric of signature $(5,n)$.
After introducing
\begin{equation}
M_{IJ} = \delta_{ab}\cV^a{}_I \cV^b{}_J  + \delta_{rs}  \cV^r{}_I \cV^s{}_J\,,
\end{equation}
the twisted selfduality equation for the tensor fields reads
\begin{equation}
\star  G_I = M_{IJ} G^J \,.
\end{equation}
Thus $G^a = \cV^a{}_I G^I$ is selfdual while $G^r = \cV^r{}_I G^I$ is anti-selfdual.

This supergravity theory describes the low energy effective theory of a type IIB superstring theory for $n=5$ and $n=21$. Type IIB string theory on a torus $T^4$, preserves all supersymmetries and gives rise to $(2,2)$ supergravity in six dimensions. By removing the two gravitini multiplets which include the 16 vector fields one obtains a truncation to $(2,0)$ supergravity coupled to $5$ tensor multiplets. The corresponding $\SO(5,5)$ metric is then the metric of the even selfdual lattice $ I\hspace{-1mm}I_{5,5}$ with split signature  
\beq \label{eta}  \eta = \left( \begin{array}{cc} 0 & \mathds{1}_5 \\ \mathds{1}_5& 0  \end{array}\right) \ . \eeq
The other possibility is type IIB string theory on K3, in which case the effective low energy theory is $(2,0)$ supergravity with 21 tensor multiplets and the $SO(5,21)$ metric is the one of the unique even selfdual lattice of signature $(5,19)$, 
\beq \eta = \left( \begin{array}{ccc} 0 & \mathds{1}_5&0 \\ \mathds{1}_5& 0  &0\\0&0 & -k\end{array}\right) \ , \eeq
with $k$ the metric of the  $\E_8\oplus \E_8$ root lattice, \ie its Cartan matrix. 

Using the gamma matrices of $\Spin(5)\cong \USp(4)$ one can express the components \(\cV^a{}_I\) as a symplectic traceless antisymmetric tensor of $\USp(4)$ with
\begin{equation}
\cV^{AB}{}_I = \cV^{[AB]}{}_I \,,\qquad \omega_{AB} \cV^{AB}{}_I = 0 \,,
\end{equation}
where $\omega_{AB}$ is the symplectic matrix with the conventions displayed in \eqref{eq:appso5gamma}.  The relation \eqref{eq:etaA} then becomes 
\begin{equation}\label{eq:etaB}
\eta_{IJ} =  \cV_{AB\,I} \cV^{AB}{}_J - \delta_{rs} \cV^r{}_I \cV^s{}_J \, ,
\end{equation}
where the $\USp(4)$ indices are raised and lowered using the symplectic matrix $\omega_{AB}$ according to \eqref{ConvRL}. One decomposes the Maurer--Cartan form  \( \dd \cV \cV^{-1}\) into its  $\mathfrak{usp}(4)$ component 
\begin{equation}\label{eq:Qconnection}
{(Q_\mu)_A}^B = \cV_{AC}{}^I \partial_\mu \cV^{BC}{}_I \,,
\end{equation}
where we use $\eta^{IJ}$ to raise and lower global $\SO(5,n)$ indices, \eg $\cV_{AB}{}^I = \omega_{AC} \omega_{BD} \eta^{IJ} \cV^{CD}{}_J$, and its coset component 
\begin{equation}
P_\mu^{AB\,r} = - \cV^{r\, I} \partial_\mu \cV^{AB}{}_I  \,.
\end{equation}
\({(Q_\mu)_A}^B\) defines the $\USp(4)$ covariant derivative \(D_\mu = \nabla_\mu + Q\).
Its action on the coset representative reads
\begin{equation}
D_\mu \cV^{AB}{}_I = \partial_\mu \cV^{AB}{}_I +2{(Q_\mu)_C}^{[A} \cV^{B]C}{}_I \,,
\end{equation}
and therefore
\begin{equation}\label{eq:covderivcoset}
D_\mu \cV^{AB}{}_I = P_\mu^{AB\, r} \cV_{r\; I} \ , \qquad D_\mu \cV^{r}{}_I = P_\mu^{AB\, r} \cV_{AB\; I}  \,. 
\end{equation}
The field strength corresponding to the connection \({(Q_\mu)_A}^B\) can be expressed in terms of \(P_\mu^{AB\,r}\), \ie
\begin{equation}\label{eq:DQ}
\left[D_\mu, D_\nu \right] X^A = \Bigl[2\partial_{[\mu} \left(Q_{\nu]}\right)_B{}^A -  2 \left(Q_{[\mu}\right)_B{}^C  \left(Q_{\nu]}\right)_C{}^A \Bigr]X^B = -2P_{[\mu \, BC\, r} P^{AC\, r}_{\nu]} X^B \,,
\end{equation}
for any \(\USp(4)\)-vector \(X^A\).

The bosonic equations of motion of the theory are given by 
\begin{align}
E_{\mu\nu} &= R_{\mu\nu} - P^{AB\,r}_\mu P^r_{AB\,\nu} - M_{IJ} G^I_{\mu\kappa\lambda} G^J_{\nu}{}^{\kappa\lambda} = 0 \,, \label{eq:einstein} \\
E^{AB\,r} &= D^\mu P^{AB\,r}_\mu - \frac23 G^{AB}_{\mu\nu\rho} G^{r\,\mu\nu\rho} = 0 \,, \label{eq:scalareom} \\
E_I &=\dd \left(\star M_{IJ} G^J \right) = 0 \,. \label{eq:3formeom}
\end{align}
The last equation is a direct consequence of the twisted selfduality equation and the Bianchi identity for the tensor fields. In practice, we shall mostly use the dressed (anti) selfdual three-form field strengths $G^{AB}  = \cV^{AB}{}_I G^I$ and $G^r = \cV^r{}_I G^I$ which transform respectively as a vector under the compact R-symmetry group $\USp(4)$ and as a vector of the flavour symmetry group $\SO(n)$. They satisfy the Bianchi identities 
\begin{equation} \label{eq:bianchi}
D G^{AB} = \delta_{rs} P^{AB\,r} \wedge G^s \,,\qquad D G^r = P^{AB\,r} \wedge G_{AB} \,. 
\end{equation}

We finally need to give the supersymmetry transformations of the fermionic fields. Under a local supersymmetry transformation the gravitini vary as
\begin{equation}\label{eq:psivariation}
\delta \psi^A_\mu = \D_\mu \epsilon^A - \frac{1}{2} G^{AB}_{\mu\nu\rho} \omega_{BC} \,\gamma^{\nu\rho} \epsilon^C \,.
\end{equation}
The supersymmetry variation of the tensorini reads
\begin{equation}\label{eq:deltachi}
\delta \chi^{Ar} =  i  P^{AB\,r}_\mu \omega_{BC} \gamma^\mu \epsilon^C + \frac{i}{12} G^r_{\mu\nu\rho} \gamma^{\mu\nu\rho} \epsilon^A \,.
\end{equation}
Notice that \(\epsilon^A\) inherits the chirality of the gravitini, \ie~\(\gamma_7 \epsilon^A  = - \epsilon^A\). 

\section{Supersymmetric solutions}\label{sec:susysolutions}

In this section we discuss necessary and sufficient conditions for the existence of a supersymmetric solution.
Here we follow the approach of \cite{Gutowski:2003rg, Cariglia:2004kk, Lam:2018jln, Cano:2018wnq} on the classification of supersymmetric solutions of  six-dimensional $(1,0)$ supergravity.

Assuming that there is at least one spinor \(\epsilon^A\) solving \(\delta \psi^A_\mu = \delta\chi^{Ar} = 0\), one can construct the following bi-linears
\begin{align}
\omega^{AB} V_\mu +  V^{AB}_\mu &= \bar\epsilon^A \gamma_\mu \epsilon^B \,, \\
\Omega^{AB}_{\mu\nu\rho} &= \bar \epsilon^A \gamma_{\mu\nu\rho} \epsilon^B \,.
\end{align}
Here, we have split \(\bar\epsilon^A \gamma_\mu \epsilon^B\) into irreducible representations of \(\USp(4)\),
\begin{equation}
V^{AB}_\mu = V^{[AB]}_\mu \,,\qquad \omega_{AB} V^{AB}_\mu = 0\,,
\end{equation}
\ie it decomposes into a singlet and a vector of $\SO(5) \cong \USp(4) / \mathds{Z}_2$,
whereas
\begin{equation}
\Omega^{AB} = \star \, \Omega^{AB} =  \Omega^{(AB)} \,,
\end{equation}
transforms in the adjoint representation.
Notice that all combinations of even rank vanish identically due to the chirality of \(\epsilon^A\).

\subsection{Algebraic constraints}

Similarly to \cite{Gutowski:2003rg} one can use Fierz identities to show that
\beq \label{eq:VdotV} \bar\epsilon^A \gamma_\mu \epsilon^B \bar\epsilon^C \gamma^\mu \epsilon^D =- \frac{1}{2} \varepsilon^{\alpha\beta\gamma\delta} \epsilon_\alpha^A \epsilon_\beta^B \epsilon_\gamma^C \epsilon_\delta^D = - \frac34 \omega^{[AB} \omega^{CD]}  \det [\epsilon] \ ,  \eeq
so that 
\begin{equation}
V_\mu V^{AB\, \mu} = 0 \ , \quad  V^{AB}_\mu V^{CD\, \mu} =  \left( \omega^{AB} \omega^{CD} + 4 \omega^{C[A} \omega^{B]D} \right)V_\mu V^\mu  \, .\end{equation}
 The reality condition on the spinor $\epsilon_\alpha^A$, 
\beq (\epsilon^*)_A^\alpha = \omega^{\alpha\beta} \omega_{AB} \epsilon^B_\beta \ , \eeq
with the symplectic form $\omega^{\alpha\beta}$ invariant under  $\USp(4) \subset \SU^*(4) \cong \Spin(1,5)$, implies that $\epsilon$ as a non-zero four by four complex matrix must be of rank 2 or rank 4, because two of its eigenvalues are the complex conjugates of the two others. We are going to see that a rank 4  $\epsilon$ is associated to supersymmetric solutions with a time-like isometry whereas a rank 2 $\epsilon$ is associated to supersymmetric solutions with a light-like isometry. Note that for a non-zero spinor $\epsilon^A$, $V^\mu$ and $V^\mu_{AB}$ are both necessarily non-zero. 
From the supersymmetry variation of the gravitini \eqref{eq:psivariation}
one obtains that 
\beq \nabla_\mu V_\nu = - \frac12 V^\sigma_{AB} G_{\mu\nu\sigma}^{AB} = - \frac12 V^\sigma_a G_{\mu\nu\sigma}^{a}  \ , \eeq
such that $V^\mu$ is a Killing vector. 

This Killing vector is time-like if $\epsilon$ is of rank 4, in which case  $V$ and $V^a$ define an orthogonal frame with $V\cdot V > 0$ and $V^a \cdot V^b = - \delta^{ab} V\cdot V $, with $\epsilon$ defining the identification of the internal $\USp(4)$ with the spin group.
On the other hand, if $\epsilon$ is of rank 2, \eqref{eq:VdotV} implies that $V\cdot V =0$.
In this case $V$ and $V^a$ are all orthogonal light-like vectors and are therefore parallel.  One can then introduce a  spacetime scalar function \(v^{AB}(x^\mu)\) such that
\begin{equation}
V^{AB}_\mu = v^{AB} V_\mu \,.
\end{equation}
In this paper we are interested in solutions generalizing already known solutions of the $(1,0)$ truncation in which the Killing vector is necessarily of null type. We shall therefore disregard the possibility of a time-like Killing vector and will concentrate on the null type in the remainder.

 It will prove convenient to introduce
\begin{equation}\label{eq:defuAB}
u^{AB} = \frac12 \left( \omega^{AB} + v^{AB} \right) \, ,
\end{equation}
such that $\bar \epsilon^A \gamma^\mu \epsilon^B  = 2 u^{AB} V^\mu$. 
Moreover, since $\epsilon$ is of rank 2, one has the additional Fierz identity 
 \beq 3 \epsilon^{[C} \bar \epsilon^A \gamma_\mu \epsilon^{B]} = 0 \  , \eeq
which implies after contraction with $\omega_{BC}$ that 
$ ( \epsilon^A - u_B{}^A \epsilon^B ) V_\mu = 0 $
and hence
\beq u_B{}^A \epsilon^B=\epsilon^A \ .   \label{eq:spinorprojectionA} \eeq
Applying this result on the definition of $u^{AB}$ shows that
\beq u_{A}{}^C u_C{}^B = u_{A}{}^B \,, 
  \eeq
which means that the matrix $u_A{}^B$ is a rank 2 projector. 
In particular 
\beq v_A{}^C v_{C}{}^B = \delta_A^B \,, \eeq
 so that $v^a$ is a  norm $2$ vector. 
One can also use the Fierz identity 
\beq \gamma^\mu \epsilon^A \bar \epsilon^B \gamma_\mu \epsilon^C = - \gamma^\mu \epsilon^C \bar \epsilon^B \gamma_\mu \epsilon^A \eeq
to show that $ \gamma^\mu \epsilon^A \bar \epsilon^B \gamma_\mu \epsilon^C$ is antisymmetric in $A,B,C$ so that for a rank 2 $\epsilon$ 
\beq \gamma^+ \epsilon^A \equiv V_\mu \gamma^\mu \epsilon^A = 0 \ . \label{eq:spinorprojectionB} \eeq
The Fierz identity 
\beq \bar \epsilon^A \gamma^\sigma \epsilon^B \bar \epsilon^C \gamma_{\mu\nu\sigma} \epsilon^D =\bar \epsilon^B \gamma^\sigma \epsilon^{C} \bar \epsilon^{D} \gamma_{\mu\nu\sigma} \epsilon^D  - 2 \bar\epsilon^A \gamma_{[\mu} \epsilon^B \bar \epsilon^C \gamma_{\nu]} \epsilon^D - 2  \bar\epsilon^C \gamma_{[\mu} \epsilon^B \bar \epsilon^A \gamma_{\nu]} \epsilon^D \eeq
implies in the rank 2 case that 
\beq V^\sigma \left( u^{AB} \Omega^{CD}_{\mu\nu\sigma} - u^{BC} \Omega^{AD}_{\mu\nu\sigma} \right) = 0 \,, \eeq
so that 
\begin{equation}\label{eq:VOmega}
\iota_V \Omega^{AB} = 0 \,.
\end{equation}
Moreover, from the rank 2 projection \eqref{eq:spinorprojectionA} it follows that
\begin{equation}\label{eq:Omegaprojection}
u_C{}^A \Omega^{CB} = \Omega^{AB} \,,
\end{equation}
so that the symmetric tensor $\Omega^{AB}$ admits only three independent components along the two dimensional subspace defined by the projection $u_B{}^A$.

Denoting the 1-form dual to \(V\) by \(e^+\), \ie \(e^+ = V_\mu \dd x^\mu\), one can always find a 1-form \(e^-\) dual to $V^\mu$, satisfying \(e^-(V) = 1\),  such that the space-time metric decomposes as
\begin{equation}\label{eq:nullbasis}
\dd s^2 = 2 e^+ e^- - \delta_{ij} e^i e^j \,,
\end{equation} 
where \(e^i\), \(i=1,\dots,4\), is an orthonormal frame of the four-dimensional space orthogonal to \(e^+\) and \(e^-\).
The orientation is fixed by \(\varepsilon^{+-ijkl} = \varepsilon^{ijkl} \). 

Finally, we also want to express the three-forms \(\Omega^{AB}\) in this basis.
Using the property \eqref{eq:VOmega} as well as their selfduality this decomposition reads
\begin{equation}
\Omega^{AB} = e^+ \wedge I^{AB} \,,
\end{equation}
where \(I^{AB} = \tfrac12 I^{AB}_{ij} e^i \wedge e^j\) are anti-selfdual two-forms with respect to the metric on the four-dimensional base space.
Using the Fierz identity 
\begin{multline} \bar \epsilon^A \gamma_{\mu\nu}{}^\lambda \epsilon^B \bar \epsilon^C \gamma_{\sigma\rho\lambda} \epsilon^D = 2 \bar \epsilon^{A)} \gamma_{[\sigma} \epsilon^{(C} \bar \epsilon^{D)} \gamma_{\rho]\mu\nu} \epsilon^{(B} - 2 \bar \epsilon^{A)} \gamma_{[\mu} \epsilon^{(C} \bar \epsilon^{D)} \gamma_{\nu]\sigma\rho} \epsilon^{(B}\\ + 4 \eta_{\sigma] [\mu} \bar \epsilon^{A)} \gamma^\lambda \epsilon^{(C} \bar \epsilon^{D)} \gamma_{\nu]\lambda[\rho} \epsilon^{(B} - 8 \eta_{\sigma][\mu} \bar \epsilon^{A)} \gamma_{\nu]} \epsilon^{(C} \bar \epsilon^{D)} \gamma_{[\rho} \epsilon^{(B} + 2 \eta_{\mu[\sigma} \eta_{\rho]\nu} \bar \epsilon^{A)} \gamma^\lambda \epsilon^{(C} \bar \epsilon^{D)} \gamma_\lambda \epsilon^{(B}  \end{multline}
and contracting it with $e_+{}^\mu e_i{}^\nu e_+{}^\sigma e_j{}^\rho$ one obtains the identity 
\beq \label{eq:hypercomplex} I^{AB}{}_i{}^k I^{CD}{}_{jk} = 4 \delta_{ij} \left( u^{AC} u^{BD} + u^{AD} u^{BC} \right)+ 4 u^{A)(C} I^{D)(B}{}_{ij} \ . \eeq
It follows that $I^{AB}$ define an almost hyper-complex structure, \ie a triplet of almost complex structures satisfying the quaternion algebra. In a basis in which $u^{12}=1$ is the only non-zero component, the triplet is defined as $\frac14( I^{11} - I^{22}), \frac i4( I^{11} +I^{22}) , - \frac i 2 I^{12} $. In particular, the $ I^{AB}$ define a complete basis of anti-selfdual forms and 
\beq \label{SelfComplete} \frac1{16} I^{AB}{}_{ij} I_{AB}{}^{kl} = \delta_{ij}^{kl} - \frac12 \varepsilon_{ij}{}^{kl} \ . \eeq

There is another useful Fierz identity 
\beq  \gamma^{\nu\sigma} \epsilon^C \bar \epsilon^A \gamma_{\mu\nu\sigma} \epsilon^B  = -8  \epsilon^{(A} \bar \epsilon^{B)}  \gamma_\mu \epsilon^{C}+ 3 \gamma_\mu \gamma_\nu  \epsilon^{(A} \bar \epsilon^{B)}  \gamma^\nu \epsilon^{C} \eeq 
that allows to show that for the rank 2 spinor
\beq \label{DiagGener} I^{AB}_{ij} \gamma^{ij} \epsilon^C = 16 u^{C(A} \epsilon^{B)} \eeq
so that the complex structures $ I^{AB}$ rotate the spinor in the representation of $\SU(2)\subset \USp(4)$. Using the relation 
\beq \{ \gamma_{\mu\nu} , \gamma^{\sigma\rho} \} = - 4 \delta_{\mu\nu}^{\sigma\rho} + \varepsilon_{\mu\nu}{}^{\sigma\rho\kappa\lambda} \gamma_{\kappa\lambda} \gamma_7 \eeq
with indices in the $\mathds{R}^4$ basis one obtains 
\beq \{ \gamma_{ij} , \gamma^{kl} \} = - 4 \delta_{ij}^{kl} - 2 \varepsilon_{ij}{}^{kl} ( 1-\gamma^- \gamma^+) \gamma_7\; ,  \eeq
which acting on $\epsilon^A$, gives
\beq \{ \gamma_{ij} , \gamma^{kl} \} \epsilon^A  = \left(- 4 \delta_{ij}^{kl} +2 \varepsilon_{ij}{}^{kl} \right) \epsilon^A \ . \eeq
It follows that 
\beq \gamma_{ij} \epsilon^A  = - \frac12 \varepsilon_{ij}{}^{kl} \gamma_{kl} \epsilon^A  = - \frac12 I_{ij}^{AB} \epsilon_B \ . \label{SelfConsepsilon} \eeq

To summarise this section, we note that the projection \eqref{eq:spinorprojectionA} halves the number of spinors to those of a $(1,0)$ theory. The additional projection \eqref{eq:spinorprojectionB} further reduces the number of supersymmetries by a half, corresponding to 1/4 BPS in the $(2,0)$ theory (\ie 1/2 BPS in the $(1,0)$ truncation).  The last constraint \eqref{SelfConsepsilon}, instead, is only satisfied for a single spinor. Therefore, we see already from the structure of the projectors that a consistent truncation to  \( (1,0)\) supergravity emerges naturally. However, this analysis is clearly not enough to guaranty that our solution is also a solution of \((1,0)\) supergravity. It will moreover be necessary that the projection matrix \({u_A}^B\) is constant and also that all fields corresponding to $(1,0)$ gravitini multiplets and (non-factorised) hyper-multiplets, notably some components of \(G^{AB}\) and \(P^{AB\,r}\), get consistently projected, too. We shall prove in the following that this is indeed the case if the solution admits four Killing spinors satisfying the same conditions \eqref{eq:spinorprojectionA} and \eqref{eq:spinorprojectionB}.

\subsection{Constraints from the Killing spinor equations}

\subsubsection*{Gravitini variation}

Let us first notice that the supersymmetry variation of the gravitini \eqref{eq:psivariation} implies
\begin{equation}
D_- \epsilon^A \equiv V^\mu D_\mu \epsilon^A = \frac12 G^{AB}_{-ij} \omega_{BC} \gamma^{ij} \epsilon^C = 0 \,,
\end{equation}
since $G^{AB}_{-ij} = V^\mu G^{AB}_{\mu ij}$ is  selfdual as a two-form on the base space, see \eqref{eq:4dselfdual}.
Consequently, we also have
\begin{equation}
\nabla_- V_\mu = D_- u^{AB} = D_- \Omega^{AB}_{\mu\nu\rho} = 0 \,.
\end{equation}
More generally, we obtain from the gravitini variation that
\begin{equation}\label{eq:DVAB}
\D_\mu \left(u^{AB} V_\nu\right) = 2 V^\kappa G^{C[A}_{\kappa\mu\nu} {u_C}^{B]} - \frac12 G^{C[A}_{\mu\kappa\lambda} {\bigl({\Omega_C}^{B]}\bigr)^{\kappa\lambda}}_\nu \,.
\end{equation}
To proceed, we split $G^{AB}$ into its component parallel and orthogonal to $v^{AB}$,
\begin{equation}\label{eq:GABdecomp}
G^{AB} = v^{AB} G + \tilde G^{AB} \,,
\end{equation}
such that $v_{AB} \tilde G^{AB} = 0$ and hence $G = \frac14 v_{AB} G^{AB}$.
Let us now also decompose \eqref{eq:DVAB} into its symplectic trace and its components parallel and orthogonal to $v^{AB}$.
Both the trace as well as the component parallel to $v^{AB}$ give
\begin{equation}\label{eq:nablaV}
\nabla_\mu V_\nu = - 2 V^\kappa G_{\kappa\mu\nu} \,,
\end{equation}
which means that $V^\mu$ is a Killing vector and
\begin{equation}
\dd V = - 4 \iota_V G \,.
\end{equation}
Furthermore, if we denote the spin-connection by $\omega$, \eqref{eq:nablaV} implies that 
\begin{equation}\label{eq:omega-}
\omega_{\mu\nu-} - 2 G_{\mu\nu-} = 0 \, . 
\end{equation}

On the other hand, the part of \eqref{eq:DVAB} orthogonal to $v^{AB}$ reads
\begin{equation}\label{eq:DuV}
\left(D_\mu u^{AB} \right) V_\nu = - V^\kappa \tilde G^{AB}_{\kappa\mu\nu} - \frac12 \tilde G^{C[A}_{\mu\kappa\lambda} {\bigl({\Omega_C}^{B]}\bigr)^{\kappa\lambda}}_\nu \,.
\end{equation}
This allows us to express the derivatives of the projector $u^{AB}$ in terms of $\tilde G^{AB}$, in components we find
\begin{equation}\begin{aligned}\label{eq:DutildeG}
D_- u^{AB} &= \tilde G^{AB}_{-ij} = 0 \,, \\
D_+ u^{AB} &= -\frac12 \tilde G_{+ij}^{C[A} \left(I_C{}^{B]}\right)^{ij} \,, \\
D_i u^{AB} &= - \tilde G^{AB}_{+-i} - \tilde G^{C[A}_{+-j} \left(I_C{}^{B]}\right)_i{}^j \,, \\
\end{aligned}\end{equation}
where the last equation is obtained by exploiting the (anti-)selfduality of $\tilde G^{AB}$ and $I^{AB}$.
On the other hand, it also follows from \eqref{eq:DuV} that
\begin{equation}
0 = \left(D_+ u^{AB} \right) V_i = \tilde G^{AB}_{+-i} + \tilde G^{C[A}_{+-j} \left(I_C{}^{B]}\right)_i{}^j \,,
\end{equation} 
and hence
\begin{equation}
D_i u^{AB} = 0 \,.
\end{equation}
Moreover, the selfduality of $\tilde G^{AB}$ and $\Omega^{AB}$ implies that
\begin{equation}
\left(D_{[\mu} u^{AB} \right) V_{\nu]} = - V^\kappa \tilde G^{AB}_{\kappa\mu\nu} \,,
\end{equation}
and therefore with the previous result we have
\begin{equation}
\tilde G^{AB}_{+-i} = 0 \,,
\end{equation}
and by selfduality also $\tilde G^{AB}_{ijk} = 0$.

In a similar fashion we obtain from the Killing spinor equation that
\begin{equation}\begin{aligned}\label{eq:DOmega}
D_{\mu} \Omega_{\kappa\lambda\rho}^{AB} &=   6 G_{\mu\sigma[\kappa} {\bigl({\Omega^{AB}}\bigr)^\sigma}_{\lambda\rho]} - 6 \tilde G^{C(A}_{\mu\sigma[\kappa} {\bigl({\Omega_C{}^{B)}}\bigr)^\sigma}_{\lambda\rho]} - 8 \tilde G^{AC}_{[\mu\kappa\lambda} V_{\rho]} {v_C}^{B} 
- 6 g_{\mu[\kappa} \tilde G^{AC}_{\lambda\rho]\sigma}  V^\sigma {v_C}^{B} \\
&= 6 G_{\mu\sigma[\kappa} {\bigl({\Omega^{AB}}\bigr)^\sigma}_{\lambda\rho]} - 6 \tilde G^{C(A}_{\mu\sigma[\kappa} {\bigl({\Omega_C{}^{B)}}\bigr)^\sigma}_{\lambda\rho]} \,, 
\end{aligned}\end{equation}
which implies that
\begin{equation}
D \Omega^{AB} = 4 V \wedge \tilde G^{AC} v_C{}^B = 0 \,.
\end{equation}
From \eqref{eq:DOmega} we also find
\begin{equation}\begin{aligned}\label{eq:DI}
D_+ I^{AB}_{ij} &= - 4 G_{+k[i} \left(I^{AB}\right)_{j]}{}^k + 4 \tilde G^{C(A}_{+k[i} \left(I_C{}^{B)}\right)_{j]}{}^k \,, \\
D_i I^{AB}_{jk} &= - 4 G_{il[j} \left(I^{AB}\right)_{k]}{}^l \,. \\
\end{aligned}\end{equation}
From the first equation in \eqref{eq:DI} we can compute
\begin{equation}\label{eq:DuI}
\left(D_+ u^{AB} \right) \left(I_B{}^C\right)_{ij} = D_+ I^{AC}_{ij} - u_B{}^A D_+ I^{BC}_{ij} = -2 \tilde G^{AB}_{+k[i} \left(I_B{}^C\right)_{j]}{}^k \,.
\end{equation}
We now use \eqref{eq:hypercomplex} to infer that
\begin{equation}\label{eq:D+uu}
\left(D_+ u^{AB}\right) u_B{}^C = - \frac{1}{48} \left(D_+ u^{AB}\right) \left(I_B{}^D\right)_{ij} \left(I_D{}^C\right)^{ij} = \frac16 \tilde G_{+ij}^{AB} \left(I_B{}^{C}\right)^{ij}\,,
\end{equation} 
which is in contradiction with \eqref{eq:DutildeG} and hence
\begin{equation}
D_+ u^{AB} = 0 \,.
\end{equation}
Moreover, using this result in \eqref{eq:DutildeG} gives
\begin{equation}
0 = \tilde G^{AB}_{+kl} \left(I_B{}^C\right)^{kl} \left(I_C{}^D\right)_{ij} = - 16 \tilde G^{AB}_{+ij} u_B{}^D - 8 \tilde G^{AB}_{+k[i} \left(I_B{}^D\right)_{j]}{}^k
\end{equation}
and since the second term vanishes due to \eqref{eq:DuI} we obtain
\begin{equation}
\tilde G^{AB}_{+ij} = 0 \,.
\end{equation}
In summary, we found that
\begin{equation}\label{eq:constantprojection}
D u^{AB} = \tilde G^{AB} = 0 \,.
\end{equation}
Note also that for a normalised $SO(5)$ vector $v_{AC} v^{BC}= \delta_A^B$ there always exists a \({\Lambda(x^\mu)_B}^A \in \USp(4)\) such that
\begin{equation}
v^{AB}(x^\mu) = {\Lambda(x^\mu)_C}^A {\Lambda(x^\mu)_D}^B v^{CD}_0 \,,
\end{equation}
for \(v^{AB}_0\) constant.
We can now use the local \(\USp(4)\) invariance of the theory to transform all fields by \(\Lambda^{-1}\).
Hence, it is always possible to choose a \(\USp(4)\) frame in which
\begin{equation} \label{UconstantFrame}
\partial_\mu u^{AB} = 0 \,.
\end{equation}
In particular, by means of \eqref{eq:constantprojection} in this frame the \(\USp(4)\) connection \({Q_A}^B\) defined in \eqref{eq:Qconnection} stabilizes \(v^{AB}\), \ie
\begin{equation}\label{eq:vstabilizer}
{Q_C}^{[A} v^{B]C} = 0 \,.
\end{equation}

Let us finally give the general form of $G$.
\eqref{eq:omega-} fixes already all its components expect for $G_{+ij}$.
To determine also these remaining components we write \eqref{eq:DI} as
\begin{equation}
D_\mu I^{AB}_{ij} = - 4 G_{\mu l[i} \left(I^{AB}\right)_{j]}{}^l \,.
\end{equation}
Using \eqref{eq:hypercomplex} and \eqref{SelfComplete} this can be solved for $G_{\mu ij}$,
\begin{equation}\label{eq:omegaG}
\left(\omega_{\mu ij} - 2 G_{\mu ij}\right)^{-} = \frac{1}{256} I_{AB\,ij} I_C{}^{A\,kl} \partial_\mu I^{CB}_{kl} + \frac14 I^{AB}_{ij} (Q_\mu)_{AB} \,,
\end{equation}
where the superscript $-$ denotes the anti-selfdual part in the indices $ij$ as a two-form on the base-space.

To write the solution more explicitly, it is convenient to introduce local coordinates \((v,u,x^m)\), \(m = 1, \dots, 4\) with $v$ the coordinate associate to the null Killing isometry $V = \partial_v$, $u$ the conjugate coordinate and $x^m$ local coordinates on the four-dimensional base space such that $\delta_{ij} e^i e^j = f^{-1} \gamma_{mn} \dd x^m \dd x^n$. The general ansatz for the metric with $V$ a null vector is then \cite{Gutowski:2003rg}
\begin{equation}\label{MetricDecompose} 
\dd s^2 = 2 f (\dd u + \beta) \left(\dd v - \tfrac12 H (\dd u+ \beta) + \omega\right) - f^{-1} \gamma_{mn} \dd x^m \dd x^n \,.
\end{equation}
Here, \(\gamma_{mn}\) is the (possibly ambipolar) metric on the four-dimensional base-space,
\(f\) and \(H\) are some functions and \(\beta = \beta_m \dd x^m \) and \(\omega = \omega_m \dd x^m\) are one-forms on the base space.
In general, all these objects depend on \(u\) and \(x^m\) but not on \(v\).
Thus, the null vielbein introduced in \eqref{eq:nullbasis} reads
\begin{equation}\begin{aligned}
e^+ = f (\dd u + \beta) \,, \qquad
e^- =  \dd v -  \tfrac12 H (\dd u+\beta) + \omega \,, \qquad
e^i = f^{-1/2} v^i \,,
\end{aligned}\end{equation}
where \(v^i = v^i_m \dd x^m\) is a vielbein of \(\gamma_{mn}\).
With these definitions it is convenient to introduce the exterior derivative on the four-dimensional base space $\tilde{d} = dx^m \partial_m$ and the derivative $D = \tilde{d} - \beta \partial_u $, as well as the Hodge star operator of the four-dimensional base space $\tilde{\star}$, defined with respect to the metric $\gamma_{mn}$.
Moreover, one can define a natural almost hyper-K\"ahler structure with respect to $\gamma_{mn}$ by introducing
\begin{equation}
J^{AB} = f I^{AB} \,,
\end{equation}
or in components $J^{AB}_{mn} = I^{AB}_{ij} v^i_m v^j_m$.

In these coordinates one can now compute the spin-connection $\omega$ and with \eqref{eq:omega-} and \eqref{eq:omegaG} one finds that $G$ takes the form \cite{Gutowski:2003rg}
\begin{equation}\begin{aligned}\label{eq:susyG}
G = &\frac14 \Bigl[ - f^{-1} e^+ \wedge e^- \wedge \bigl(D f - f \dot \beta \bigr)  - f e^- \wedge D \beta   \\ 
&\qquad +  \tilde \star \bigl(D f^{-1} + f^{-1} \dot \beta \bigr) +  e^+ \wedge \left( \left(D \omega\right)^{-} - f^{-2} \psi \right) \Bigr] \,,
\end{aligned}\end{equation}
where $\psi$ is an anti-selfdual two-form on the base, defined by
\begin{equation}
\psi =   \left( \frac{1}{128}J_{AC}^{mn} \partial_u J_B{}^C{}_{mn} + \frac12 (Q_u)_{AB} \right) J^{AB} \,.
\end{equation}

These results show for the gravitational multiplet sector that every supersymmetric configuration is at the same time also a supersymmetric configuration of a $(1,0)$ theory, in the sense that the tensor fields of the gravitino multiplets vanish. While $G$ corresponds to the selfdual three-form in the $(1,0)$ gravity multiplet, the remaining four components $\tilde G^{AB}$, which would be part of a $(1,0)$ gravitino multiplet, are consistently projected out.

\subsubsection*{Tensorini variation}

We now carry over with the analysis of the tensor multiplet sector. 
Contracting the spin-1/2 variation \eqref{eq:deltachi} with \(\bar\epsilon^B\) yields 
\begin{equation}
24 V^\mu P_\mu^{AC\,r} {u_C}^B + \left(\Omega^{AB}\right)^{\mu\nu\rho} G^r_{\mu\nu\rho} = 0 \,.
\end{equation}
Both terms transform in different representations of $\USp(4)$ and therefore must vanish independently,
\begin{equation}\label{eq:p-}
P_-^{AB\,r} \equiv V^\mu P_\mu^{AB\,r} = 0 \,.
\end{equation}
On the other hand, contracting \eqref{eq:deltachi} with \(\bar\epsilon^B \gamma_{\mu\nu}\) gives 
\begin{equation}\label{eq:deltachiA}
8 P_{[\mu}^{AC \,r} {u_C}^B V_{\nu]} + 2 P_{\kappa}^{AC \,r} {\left({\Omega_C}^B\right)_{\mu\nu}}^\kappa + 2 G^r_{\kappa\lambda[\mu} {\left(\Omega^{AB}\right)_{\nu]}}^{\kappa\lambda} - 4 u^{AB}G^r_{\mu\nu\kappa} V^\kappa
 = 0 \,.
\end{equation}
As before for $G^{AB}$, we split $P^{AB\,r}$ according to
\begin{equation}
P^{AB} = v^{AB} P^r + \tilde P^{AB\,r} \,,
\end{equation}
such that $v_{AB} \tilde P^{AB\,r} = 0$ and hence $P^r = \frac14 v_{AB} P^{AB\,r}$.
The symplectic trace of \eqref{eq:deltachiA} and its component parallel to $v^{AB}$ read
\begin{equation}\label{eq:iVGr}
V \wedge P^r = - \iota_V G^r \,.
\end{equation}
Therefore, $G^r$ takes the general form
\begin{equation}\label{eq:Gr}
G^r = (1- \star) \, e^+ \wedge e^- \wedge P^r  + e^+ \wedge F^r_{\scalebox{0.5}{(+)}} \,,
\end{equation}
where $F^r_{\scalebox{0.5}{(+)}} $ are arbitrary selfdual two-forms on the four-dimensional base space, see also \eqref{eq:Gdecomp}.
From the part of \eqref{eq:deltachiA} which is orthogonal to $v^{AB}$ we infer that
\begin{equation}\label{eq:tildePprojection}
\tilde P^{AC\,r}_i {u_C}^{B} = \frac12  \tilde P^{AC\,r}_j {\bigl({I_C}^{B}\bigr)_i}^j \,.
\end{equation}
Let us finally notice that there is an integrability condition on \eqref{eq:constantprojection} which reads
\begin{equation}\label{eq:tensorhyperorth}
\left[ D_\mu, D_\nu \right] u^{AB} = - 4 \tilde P^{AB\,r}_{[\mu} P^r_{\nu]} = 0 \,.
\end{equation}

\subsection{Number of independent Killing spinors}\label{sec:numberofSUSYs}

In the previous section we have determined necessary conditions for the existence of a supersymmetric configuration of $ (2,0)$ supergravity with a null isometry.
In particular, they resemble very closely the conditions for a supersymmetric configuration of $(1,0)$ theories.
It remains to verify that these conditions are also sufficient.
In the following we consider spinors $\epsilon^A$ satisfying the conditions \eqref{eq:spinorprojectionA} and \eqref{eq:spinorprojectionB}, each reducing the number of supersymmetries by a factor of 1/2, such that only four of the original sixteen supercharges are preserved.
The presence of hypermultiplet scalars $\tilde P^{AB\,r}$ will require another constraint on $\epsilon^A$. 

To determine if the previously determined conditions on \(P^{AB\,r}\) and \(G^r\) are sufficient for the existence of a solution of \(\delta\chi = 0\) we insert \eqref{eq:Gr} back into \eqref{eq:deltachi}.
Using \eqref{eq:spinorprojectionB} and \eqref{eq:p-} this gives
\begin{equation}\begin{aligned}
\delta \chi^{A\,r} &=  i P_i^{AB\,r} \omega_{BC} \gamma^i \epsilon^C +i G^r_{+-i} \gamma^i \epsilon^A 
= i \tilde P_i^{AB\,r} \omega_{BC} \gamma^i \epsilon^C \,.
\end{aligned}\end{equation}
For a single supersymmetry parameter, $\epsilon^A$ satisfies 
\begin{equation}\label{eq:spinorprojectionC}
\gamma^i \epsilon^A = \frac16 \left(I^{AB}\right)_j{}^i \gamma^j \epsilon_B \,,
\end{equation}
so that the constraint $ \tilde P_i^{AB\,r} \omega_{BC} \gamma^i \epsilon^C=0$ is automatically satisfied. Notice that \eqref{eq:spinorprojectionC} is equivalent to \eqref{SelfConsepsilon} and therefore always satisfied by the spinor from which $I^{AB}$ is defined.
 On the contrary, if one has four preserved supercharges, with the four linearly independent spinors $\epsilon^A$ satisfying  \eqref{eq:spinorprojectionA} and \eqref{eq:spinorprojectionB}, then one gets $ \tilde P_i^{AB\,r}=0$. In principle there may  exist intermediary solutions with only two or three independent supersymmetries and the $ \tilde P_i^{AB\,r}$ are further constrained accordingly.

After inserting \eqref{eq:constantprojection} into \eqref{eq:psivariation} the gravitino variation $\delta \psi^A_\mu = 0$  reduces to 
\begin{equation}
D_\mu \epsilon^A + \frac12 G_{\mu\nu\rho} \gamma^{\nu\rho} \epsilon^A = 0 \,.
\end{equation}
Following \cite{Gutowski:2003rg, Cariglia:2004kk}, this equation can be written as
\begin{equation}
\partial_\mu \epsilon^A - \frac14 \left(\omega_{\mu\nu\rho} - 2 G_{\mu\nu\rho}\right) \gamma^{\nu\rho} \epsilon^A + \left(Q_\mu\right)_B{}^A \epsilon^B = 0 \,,
\end{equation}
and using \eqref{eq:spinorprojectionA} and \eqref{eq:omega-} it becomes
\begin{equation}
\partial_\mu \epsilon^A - \frac14 \left(\omega_{\mu ij} - 2 G_{\mu ij}\right) \gamma^{ij} \epsilon^A + \left(Q_\mu\right)_B{}^A  \epsilon^B= 0 \,.
\end{equation}
To proceed, we use \eqref{eq:omegaG} and obtain
\begin{equation}
\partial_\mu \epsilon^A - \frac14 \left[\frac{1}{256} I_{BC\,ij} I_D{}^{B\,kl} \partial_\mu I^{DC}_{kl} + \frac14 I^{BC}_{ij} (Q_\mu)_{BC}\right]\gamma^{ij} \epsilon^A + \left(Q_\mu\right)_B{}^A \epsilon^B = 0  \,.
\end{equation}
As for the $(1,0)$ theory  \cite{Gutowski:2003rg}, the solution to \eqref{eq:hypercomplex} can always be chosen up to an $SU(2)\subset SO(4)$ local transformation on the frame $e^i$ such that the $I_{ij}^{AB}$ are canonical constant coefficients, so that $\partial_\mu I^{AB}_{ij} = 0$, and one gets 
\begin{equation}
\partial_\mu \epsilon^A - \frac1{16}  I^{BC}_{ij} (Q_\mu)_{BC} \gamma^{ij} \epsilon^A + \left(Q_\mu\right)_B{}^A \epsilon^B = 0  \,.
\end{equation}
Once again this is automatically integrable in the two extreme cases discussed above. Either one has only one supercharge satisfying \eqref{DiagGener}, and after choosing a $\USp(4)$ gauge such that \eqref{UconstantFrame} one obtains that this equation reduces to $\partial_\mu \epsilon^A=0$. If we have instead four independent supercharges, then $\tilde{P}_i^{AB}=0$ and thus \eqref{eq:DQ} and \eqref{eq:tensorhyperorth} imply
\beq d Q_B{}^A - Q_B{}^C \wedge  Q_C{}^B = 0 \ . \eeq
Consequently, we can find a $\USp(4)$ gauge such that $Q_B{}^A=0$ and obtain $\partial_\mu \epsilon^A = 0$ again.
For the intermediary case with two or three Killing spinors one will get further constraints on $Q_B{}^A$. 

\subsection{Equations of motion}
\label{EMsection} 
Let us finally discuss the equations of motion and inspect under which conditions a supersymmetric configuration, such that the supersymmetry variations of the fermionic fields vanish, is also a  solution of the equations of motion.
It was found in \cite{Gutowski:2003rg, Cariglia:2004kk, Lam:2018jln, Cano:2018wnq} that a supersymmetric configuration of a $ (1,0)$ theory is automatically also a solution of its equations of motion if moreover the three-form Bianchi identities are satisfied, as well as the $++$ component of the Einstein equations.  All remaining equations of motion are already implied by the Killing spinor equations.
As we show in Appendix~\ref{app:integrability} the same holds true for the $ (2,0)$ theories.
Under the previously determined conditions, the Einstein equations \eqref{eq:einstein} reduce to
\begin{equation}
R_{\mu\nu} - 4 G_{\mu\kappa\lambda} {G_\nu}^{\kappa\lambda} -  G^r_{\mu\kappa\lambda} {G_\nu}^{\kappa\lambda}{}_{r} - 4 P^r_\mu P_{\nu\, r}  - \tilde P^{AB\,r}_\mu \tilde P_{\nu\, AB\, r}   = 0 \,,
\end{equation}
while the scalar equations of motion \eqref{eq:scalareom} split into
\begin{equation}\label{eq:10scalareom}
D^\mu P^{r}_\mu =  \frac23 G_{\mu\nu\rho} G^{r\,\mu\nu\rho} \,,\qquad D^\mu \tilde P^{AB\,r}_\mu = 0 \,,
\end{equation}
and the Bianchi identities \eqref{eq:bianchi} give
\begin{equation}\label{eq:10bianchi}
d G = P^r \wedge G^r \,,\qquad D G^r = 4 P^r \wedge G \,,
\end{equation}
as well as
\begin{equation}\label{eq:10bianchiconstraint}
\tilde P^{AB\,r} \wedge G^r = 0 \,.
\end{equation}
These equations resemble closely the corresponding equations of $ (1,0)$ supergravity with an additional constraint \eqref{eq:10bianchiconstraint} that would trivially be satisfied in $ (1,0)$ supergravity. However, $P^r $ and $P^{AB\,r}$ are still momenta of the full coset space  $SO(5,n) / (SO(5) \times SO(n))$, and we must explicitly decompose this coset space into a tensor multiplet moduli space $\SO(1,n_{\scalebox{0.5}{T}}) / \SO(n_{\scalebox{0.5}{T}})$ and a quaternionic K\"ahler coset space $\SO(4,n_{\scalebox{0.5}{H}}) / (\SO(4) \times \SO(n_{\scalebox{0.5}{H}}))$ in order to understand the solution in $(1,0)$ supergravity. 

To proceed, we make use of the fact one can always write the coset representative $\cV$ as 
\begin{equation}\label{eq:cosetsplit}
\cV = \cVT \cVH \,,
\quad\text{with}\quad
\cVT \in \SO(1,n) \quad\text{and}\quad \cVH \in \SO(4,n) \,.
\end{equation}
Here, $\cVH$ carries the right rigid $\SO(5,n)$ index of $\cV$ and a left local $\SO(5) \times \SO(n)$ vector index, \ie it takes the form
\begin{equation}
\cVH = (\cVH^{\underline{AB}}{}_I, \cVH^{\underline{r}}{}_I)^T \,,
\end{equation}
while $\cVT$ has only $SO(5)\times SO(n)$ indices, so its components are given by
\begin{equation}
\cVT = \begin{pmatrix} \cVT^{AB}{}_{\underline{CD} }&  \cVT^{AB}{}_{\underline s} \\ \cVT^r{}_{\underline{CD}} & \cVT^r{}_{\underline s} \end{pmatrix} \,.
\end{equation}
The underlined indices are associated to the ambiguity in the split of $\cV$, which we fix partially by imposing the following conditions on $\cVT$ and $\cVH$. According to the previously introduced notation we take the $\USp(4)$ frame in which $v^{AB}$ is constant and decompose $\cVT$ as
\begin{equation}
\cVT^{AB}{}_{\underline{CD}} = v^{AB} \cVT_{\underline{CD}} + \tcVT{}^{AB}{}_{\underline{CD}} \,,
\end{equation}
with\begin{equation}
\cVT_{\underline{AB}} = \tfrac12 v_{\underline{AB}} \cV_T \,,\qquad \tcVT{}^{AB}{}_{\underline{CD}} =  \delta^{[A}_{\underline{C}} \delta^{B]}_{\underline{D}} - \tfrac{1}{4} \omega^{AB} \omega_{\underline{CD}} - \tfrac14 v^{AB} v_{\underline{CD}}  \,,
\end{equation}
where the components of $\omega^{\underline{AB}}$ and $v^{\underline{AB}}$ are the same as those of  $\omega^{{AB}}$ and $v^{{AB}}$, and we only keep the underlined indices to recall that they are associated to the spurious $SO(5) \times SO(n)$ that we have introduced in the splitting \eqref{eq:cosetsplit}. The remaining components of $\cVT$ satisfy
\begin{equation}
\cVT^{AB}{}_{\underline r} = v^{AB} \cVT_{\underline r} \,,\qquad \cVT^r{}_{\underline {AB}} = \tfrac12 v_{\underline{AB}} \cVT^r \,, 
\end{equation}
while $\cVT^r{}_{\underline s}$ is unconstrained.
Notice, that $\cVT \in \SO(1,n)$ implies
\begin{equation}\label{eq:SO1nproperties}
4 \cVT \cVT - \cVT^r \cVT_r = 1\,,\qquad 4 \cVT \cVT_{\underline{r}} - \cVT^r \cVT_{r \underline{r}} = 0 \,,\qquad  4 \cVT_{\underline r} \cVT_{\underline s} - \cVT^r{}_{\underline r}  \cVT_{r \underline s} = - \delta_{\underline {rs}} \; . 
\end{equation}
Similarly, we decompose $\cVH$ according to
\begin{equation}
\cVH^{\underline{AB}}{}_I = \tfrac{1}{4} v^{\underline{AB}} \; v_I + \tcVH{}^{\underline{AB}}{}_I \,, \qquad v_{\underline{AB}} \tcVH{}^{\underline{AB}}{}_I = 0 \,,
\end{equation}
such that $v_I$ is a constant $\SO(5,n)$ vector of norm $2$. In particular, this decomposition implies
\begin{equation} \label{etaHyper} 
\tfrac{1}{4} v_I v_J + \tcVH^{\underline{AB}}{}_I \tcVH{}_{\underline{AB}\, J} -  \cVH^{\underline{r}}{}_I {\cVH}{}_{\underline{r}\, J} = \eta_{IJ} \,.
\end{equation}
Notice, that $\cVH$ and $\cVT$ are defined only up to an arbitrary local $SO(n)$ transformation acting on the underlined indices.

Following this decomposition, we can now compute
\begin{equation}
2 Q_C{}^{[A} v^{B]C} = - 4\delta^{A}_{\underline{C}} \delta^{B}_{\underline{D}}  \PH^{\underline{CD}\,\underline r} \cVT_{\underline r} \,,
\end{equation}
where $\PH^{\underline {AB}\,\underline r}$ denotes the respective component of the Maurer--Cartan form of  $\cVH$. 
Hence, $D v^{AB} = 0$ implies 
\begin{equation}\label{eq:VTPH}
\cVT_{\underline r} \PH^{\underline{AB}\,\underline r} = 0 \,. 
\end{equation}
Using this result we determine the various other components of the Maurer-Cartan form.
For the $\USp(4)$-part of the composite connection \eqref{eq:Qconnection} we find
\begin{equation}
Q_A{}^B = \delta_A^{\underline A} \delta_{\underline B}^B \QH{}_{\underline A}{}^{\underline B} \,,
\end{equation}
which we shall   abbreviate as  $\QH{}_A{}^B$,
while the $\SO(n)$ part of the connection reads
\begin{equation}
Q^{rs} = \QT^{rs} + \cVT^r{}_{\underline r} \QH^{\underline {rs}} \cVT^s{}_{\underline s} \,.
\end{equation}
For $P^{AB\,r}$, on the other hand, we find
\begin{equation}\label{eq:decompP}
P^r = \PT^r - \cVT^r{}_{\underline s} \QH^{\underline {st}} \cVT_{\underline t} \,,
\end{equation}
and
\begin{equation}\label{eq:decomptildeP}
\tilde P^{AB\,r} =  \cVT^r{}_{\underline r}   \delta^{A}_{\underline{C}} \delta^{B}_{\underline{D}}   \PH^{\underline{CD}\,\underline r}  \,.
\end{equation}
One can show that \eqref{eq:decompP} and \eqref{eq:decomptildeP} together with \eqref{eq:VTPH} satisfy \eqref{eq:tensorhyperorth}.
Moreover, we find
\begin{equation}
D_\mu \tilde P^{AB\,r}_\nu = \cVT^r{}_{\underline r} \delta^{A}_{\underline{C}} \delta^{B}_{\underline{D}} \,  D_{\mu}^{\scalebox{0.5}{H}}   P_{\scalebox{0.5}{H}\,\nu}^{\underline{CD}\,\underline r} \,,
\end{equation}
where $\dH$ denotes the covariant derivative with respect to only $\QH$.
The second equation in \eqref{eq:10scalareom} is thus equivalent to
\begin{equation}
 \dH^\mu  P_{\scalebox{0.5}{H}\,\mu}^{\underline{AB}\,\underline r} = 0 \,,
\end{equation}
and indeed describes $(1,0)$ hypermultiplet scalar fields parametrising the quaternionic K\"ahler manifold $\SO(4,n) / (\SO(4) \times \SO(n))$.
However, due to the mixing with $\QH$ in \eqref{eq:decompP}, $P^r$ does not directly correspond to scalars on a coset space of the form $\SO(1, n) / \SO(n)$.
In particular the first equation of \eqref{eq:10scalareom} depends non-trivially on both $\cVT$ and $\cVH$, which is not the case in a genuine $(1,0)$ theory.
Instead, the scalar geometry is described by a fibration of $\SO(1, n) / \SO(n)$ over $\SO(4,n) / (\SO(4) \times \SO(n))$.

To continue the discussion of the tensor multiplet sector, let us furthermore rewrite the constraint  $\tilde{G}^{AB}=0$ as
\beq \label{GAB} \tilde{G}^{AB} = \tcVT{}^{AB}{}_{\underline{CD}}  \cVH^{\underline{CD}}{}_I G^I =   \delta^{A}_{\underline{C}} \delta^{B}_{\underline{D}}  \tcVH{}^{\underline{CD}}{}_I G^I  = 0 \; .  \eeq
It follows using \eqref{etaHyper} that 
\beq -  \cVH^{\underline{r}\, I} \cVH_{\underline{r}\, J} G^J = ( \delta^I_J - \tfrac14 v^I v_J ) G^J \;  . \eeq
Therefore, $-  \cVH^{\underline{r}\, I} \cVH_{\underline{r}\, J} $ acts on $G^I$ as the constant projection onto the subspace orthogonal to the constant vector $v_I$. Moreover, because $\cVH^{\underline{r}}{}_I \cVH_{\underline{r}\, J} $ is by construction a positive definite matrix, one can define this projection as a manifestly positive constant matrix $v^{\underline{r}}{}_I v_{\underline{r} J}$. Moreover, one checks that $\cVH^{\underline{r}}{}_I v_{\underline{s} I}$ acts as an $SO(n)$ rotation on $\cVH_{\underline{r}\, J} G^J$, so that one can use the local $SO(n)$ invariance to choose $ \cVH_{\underline{r}\, J} $ such that  $ \cVH^{\underline{r}}{}_{J} G^J = v^{\underline{r}}{}_J G^J$. The set of matrices $(v_I , v^{\underline{r}}{}_I)$ satisfies by construction 
\beq \bigl( \tfrac{1}{4} v_I v_J - v^{\underline{r}}{}_I v_{\underline{r} \, J}   \bigr) G^J = \eta_{IJ} G^J \,, \eeq
and one chooses the constant matrices $v^{\underline{r}}{}_I$ of lowest possible rank $n_{\scalebox{0.5}{T}}$ such that this identity holds for all $G^I$ at all points in spacetime. By definition, $n_{\scalebox{0.5}{T}}$ is the dimension of the span of $\cVH^{\underline{r}}{}_I  G^{I}$ in $\mathds{R}^n$. The set of matrices $(v_I , v^{\underline{r}}{}_I)$ then define a constant metric of signature $(1,n_{\scalebox{0.5}{T}})$ 
\beq  \label{etaT} \eta^{\scalebox{0.5}{T}}_{IJ} \equiv  \tfrac{1}{4} v_I v_J - v^{\underline{r}}{}_I v_{\underline{r} \, J}   \;  , \eeq 
which acts trivially on $G^I$, \ie
\begin{equation}
\eta^{\scalebox{0.5}{T}}_{IJ}  G^J =  \eta_{IJ}  G^J \, . 
\end{equation}
$\eta^{\scalebox{0.5}{T}}_{I}{}^{J} = \eta^{\scalebox{0.5}{T}}_{IK} \eta^{JK}  $ then defines a projector onto a subspace of dimension $1+ n_{\scalebox{0.5}{T}}$.  One can verify that $v^I$ and $ v^{\underline{r}}{}_I $ are orthogonal in this subspace,  so
\begin{equation}\label{eq:vinvers}
v^{\underline r}{}_I v^I = 0 \,,\qquad  v^{\underline{r}}{}_I  v^{\underline{s}\,I}= - \delta^{\underline{rs}}_{\scalebox{0.5}{T}} \,,
\end{equation}
where $\delta^{\underline{rs}}_{\scalebox{0.5}{T}}$ is a rank $n_{\scalebox{0.5}{T}}$ projector that satisfies $G^{\underline r} = \delta^{\underline{rs}}_{\scalebox{0.5}{T}} G_{\underline s}$.
Therefore, we can define 
\beq \label{eq:newcVT} \cVT_I \equiv  \cVT v_I + \cVT_{\underline{r}}v^{\underline{r}}{}_I \; , \qquad  \cVT^r{}_I \equiv  \cVT^r v_I + \cVT^r{}_{\underline{s}}v^{\underline{s}}{}_I \; , \eeq
satisfying 
\beq 4 \cVT_I \cVT_J - \cVT^r{}_I  \cVT_{r \, J}  =  \eta^{\scalebox{0.5}{T}}_{IJ}  \, . \eeq 
By construction we have  
\beq G = \cVT_I G^I \; , \qquad G^r = \cVT^r{}_I G^I \; . \eeq
and hence we would like to identify $(\cVT_I, \cVT^r{}_I)$ with the tensor multiplet coset representative. However, this would require  $\PT^r = - \eta^{IJ}_{\scalebox{0.5}{T}} \cVT^r{}_I d \cVT^I$ which would only be valid if $  \delta^{\underline{rs}}_{\scalebox{0.5}{T}}d \cVT_{\underline{s}} =d \cVT^{\underline{r}} $, but this does not need to be true in general. 

This will be true if we suppose the additional constraint
\begin{equation}\label{eq:10condition}
\cVT_{\underline r} \QH^{\underline {rs}} = 0 \,,
\end{equation}
which is together with \eqref{eq:VTPH} equivalent to
\begin{equation}
\cVT_{\underline r} \, \partial_\mu \cVH^{\underline r}{}_I = 0 \,.
\end{equation}
This conditions implies that the tensor and the hypermultiplets are locally defined in orthogonal subspaces. In particular, one obtains then that they decouple and 
$P^r = \PT^r$
as well as
$D_\mu P^r_\nu = \DT{}_\mu P_{\scalebox{0.5}{T}\,\nu}^r$.
Hence, the first equation in \eqref{eq:10scalareom} becomes
\begin{equation}
\DT^\mu P_{\scalebox{0.5}{T}\,\mu}^r = \frac23 G_{\mu\nu\rho} G^{r\,\mu\nu\rho} \,.
\end{equation}
Moreover, it now follows from $\left(\delta^{\underline {rs}} - \delta^{\underline {rs}}_{\scalebox{0.5}{T}}\right) G_{\underline s} = 0$ in combination with the supersymmetry conditions \eqref{eq:susyG} and \eqref{eq:Gr} that $\left(\delta^{\underline {rs}} - \delta^{\underline {rs}}_{\scalebox{0.5}{T}}\right) D \cVT_{\underline s} \propto \left(\delta^{\underline {rs}} - \delta^{\underline {rs}}_{\scalebox{0.5}{T}}\right) \cVT_{\underline s}$.
The only solution to this relation is $\left(\delta^{\underline {rs}} - \delta^{\underline {rs}}_{\scalebox{0.5}{T}}\right) \cVT_{\underline s} = 0$ and hence
\begin{equation}
\PT^r = - 4 \cVT^r d \cVT + \delta^{\underline {st}}_{\scalebox{0.5}{T}} \cVT^r{}_{\underline s} d \cVT_{\underline t} =  - \eta^{IJ}_{\scalebox{0.5}{T}} \cVT^r{}_I \d \cVT_J \, . 
\end{equation}
Therefore, under the assumption that \eqref{eq:10condition} is satisfied, we can indeed identify $(\cVT_I, \cVT^r{}_I)$ with the tensor multiplet coset representative.
It is then convenient to introduce 
\begin{equation}
M^{\scalebox{0.5}{T}}_{IJ} = 4 \cVT_I \cVT_J + \cVT^r{}_I \cVT^r{}_J \,,
\end{equation}
so
\begin{equation}
\star G_I = M^{\scalebox{0.5}{T}}_{IJ} G^J \,,
\end{equation}
and the Einstein equations take the form
\begin{equation}
R_{\mu\nu} - 4 \delta_{rs} \PT^r{}_\mu \PT^s{}_\nu - M^{\scalebox{0.5}{T}}_{IJ} G^I{}_{\mu\kappa\lambda} G^J{}_{\nu}{}^{\kappa\lambda} - \delta_{\underline {rs}} \PH^{\underline {AB}\,\underline r}{}_\mu \PH{}^{\underline r}_{\underline{AB}\,\nu} = 0 \,.
\end{equation}
We have thus shown that \eqref{eq:10condition} is a sufficient condition for a supersymmetric solution of $(2,0)$ supergravity to satisfy the equations of motions of $(1,0)$ supergravity.

We shall see in the next section that in the special case of $\tilde P_i^{AB\,r} = 0$, \ie when the hypermultiplets only depend on the coordinate $u$, one can choose a gauge in which \eqref{eq:10condition} is indeed satisfied. This is also the case if the the moduli space factorises completely, \ie if
\beq   \delta^{\underline{rs}}_{\scalebox{0.5}{T}}d \cVT_{\underline{s}} =d \cVT^{\underline{r}}  \,, \qquad  \delta^{\underline{rs}}_{\scalebox{0.5}{T}}d \cVH_{\underline{s} I} =0 \; ,
 \eeq
in which case one can simply take $v^{\underline{r}}{}_I =  \delta^{\underline{rs}}_{\scalebox{0.5}{T}} \cVH_{\underline{s} I}$ which is constant.

\subsection{Solutions with four supercharges}

We found in Section~\ref{sec:numberofSUSYs} that a generic solution with a light-like isometry preserves only one supersymmetry.
In the following, let us however focus on solutions preserving four independent supersymmetries, \ie~1/4 BPS solutions, which are only constrained by \eqref{eq:spinorprojectionA} and \eqref{eq:spinorprojectionB}.
From the discussion of the tensorini variation we know that in this case
\begin{equation}\label{eq:tildePi}
\tilde P^{AB\,r}_i = 0 \,,
\end{equation}
and the only component of $\tilde P^{AB\,r}$ which can be possibly non-vanishing is $\tilde P^{AB\,r}_+$.
This component of $P^{AB\,r}$ is projected out of \eqref{eq:deltachi} by \eqref{eq:spinorprojectionB} and is therefore unconstraint by the supersymmetry variations.
 Notice that \eqref{eq:tildePi} is equivalent to $\PH^{\underline{AB}\, \underline r}{}_i=0$ according to \eqref{eq:decomptildeP}. 
Moreover, as discussed at the end of Section~\ref{sec:numberofSUSYs}, \eqref{eq:tildePi} implies that $Q_A{}^B$ is locally pure gauge and hence can be chosen to vanish.

However, there is an integrability condition on \eqref{eq:tildePi} that can be used to further constrain $\tilde P^{AB\,r}_+$.
For this purpose we use the explicit split  \(\cV^{AB}{}_I = v^{AB} \cV_I + \tilde{\cV}^{AB}{}_I \)  to rewrite \eqref{eq:p-} and \eqref{eq:tildePi} as
\begin{equation}\label{eq:cosetderivs}
D_- \cV^{AB}{}_I = D_i \tilde\cV^{AB}{}_I = 0 \,,
\end{equation}
which follows from \eqref{eq:covderivcoset} and \eqref{eq:constantprojection}.
Again, $D_+ \tilde \cV^{AB}{}_I$  is a priori unconstraint.
The Frobenius integrability condition for \eqref{eq:cosetderivs} reads
\begin{equation}\label{eq:integrabilityV}
\left[e_i + Q_i, e_j + Q_j\right] \tilde\cV^{AB}{}_I = P^r_{[i\, CD} P_{j]}^{r}{}^{D[A} \tilde\cV^{B]C}{}_I + \left[e_i, e_j\right]^\mu D_\mu \tilde\cV^{AB}{}_I \,,
\end{equation}
where $e_i + Q_i $ is the vector field $e_i$ acting as a Lie derivative plus the $\USp(4)$ connection along $e_i$. 
Firstly, we notice that \eqref{eq:tildePi} implies
\begin{equation}
P^r_{[i\, AC} P^{r\,CB}_{j]} = - P_{[i} P_{j]} \delta_A^B = 0 \,.
\end{equation}
Moreover, the commutator between two frame vector fields \(e_i{}^\mu \) can be expressed in terms of the spin-connection \(\omega\), so \eqref{eq:integrabilityV} reduces to
\begin{equation}
\left[e_i + Q_i, e_j + Q_j\right]  \tilde\cV^{AB}{}_I = 2 {\omega_{ij}}^+ D_{+} \tilde\cV^{AB}{}_I \,.
\end{equation}
Hence, $\tilde P^{AB\,r}_+$ can only be non-trivial if $\omega_{ij}{}^+ = 0$.

For the general supersymmetric metric \eqref{MetricDecompose} this component of the spin-connection reads \cite{Gutowski:2003rg}
\begin{equation}
\omega_{ij-} = - \frac12 f (D \beta)_{ij} \,,
\end{equation}
where $(D\beta)_{ij}$ was introduced below \eqref{MetricDecompose}. $f^{-1} \gamma_{ij}$ is a regular metric on the base space, and the function $f$ can only vanish on measure zero surfaces interpreted as evanescent ergosurfaces \cite{Niehoff:2016gbi}. The condition $\omega_{ij}{}^+ = 0$ therefore requires $D\beta=0$ by continuity.  
One can interpret $\beta(x,u)$ as a connection over the four-dimensional base of a Virasoro group acting on the circle parametrized by $u$.
$D\beta$ is then its field strength and  for $D\beta=0$, $\beta$ is a flat connexion which is locally pure gauge and can be written as 
\beq \beta(x,u) = \frac{ \partial_{m}  \alpha(x,u)}{1+\partial_u \alpha(x,u)} dx^m\,. \eeq
Thus, such a flat connection $\beta$ can always be reabsorbed by a change of coordinate $u\rightarrow u - \alpha(x,u)$ and a redefinition of the function $f$.

To summarize, we can distinguish two branches of solutions.
The first is characterised by $D\beta \neq 0$ which in turn enforces
\begin{equation}
\tilde P^{AB\,r}_+ = 0 \,,
\end{equation}
so there are no $(1,0)$ hypermultiplets.
On the second branch we have $\beta = 0$ and $\tilde P^{AB\,r}_+$ is unconstrained.
The corresponding equation of motion \eqref{eq:10scalareom} reads
\begin{equation}
D^\mu \tilde P^{AB\,r}_\mu = D_{-} \tilde P^{AB\,r}_+ = 0 \,,
\end{equation}
and is identically satisfied.

Let us finally discuss the implications of \eqref{eq:tildePi} on the split of the coset representative \eqref{eq:cosetsplit}.
In particular, since $\cVT^r{}_{\underline s}$ is an invertible matrix, we find from \eqref{eq:decomptildeP} that
$\PHi^{\underline {AB}\,\underline r}{} = 0$
and therefore
\begin{equation}
\PH^{\underline {AB}\,\underline r} \wedge \PH^{\underline {CD}\,\underline s} = 0 \,.
\end{equation}
Consequently, the curvature of $\QH$ vanishes and we can choose a gauge such that
\begin{equation}\label{eq:noQH}
\QH = 0 \,,
\end{equation}
and in which $\cVH$ is a function of $u$ only. This condition implies \eqref{eq:10condition}, therefore, according to the discussion of the preceding section, every solution with four supercharges is a solution of a $(1,0)$ theory, where the hypermultiplet scalars are arbitrary functions of $u$, satisfying pointwise the algebraic constraints  \eqref{eq:VTPH} and \eqref{GAB}.



\section{Microstate geometries}\label{sec:microstates}
 Let us now come back to the discussion of microstate geometries and their dependence on the moduli parametrizing $\SO(5,n)/ (\SO(5) \times \SO(n))$, for $n=5$ or $21$. It is convenient for the microstate geometry interpretation to write the metric \eqref{MetricDecompose} as
 \begin{equation} \label{MetricForm} 
\dd s^2 = - f H \bigl( du + \beta - H^{-1}( dv + \omega ) \bigr)^2  + \frac{f}{H} (dv + \omega)^2   - f^{-1} \gamma_{mn} \dd x^m \dd x^n \,.
\end{equation}
 such that one can identify $u$ as the coordinate of a circle fibered over a five-dimensional pseudo-Riemannian spacetime. The microstate geometry generically depends non-trivially on the circle coordinate $u$, but asymptotically the leading contributions to the metric are constant on the circle, and the metric approaches the one of a black hole solution for which $\partial_u$ is an isometry. The metric field \eqref{MetricForm} of the black hole solution can therefore be decomposed into an additional dilaton $R_y = f H$, a vector field $A^3 =  \beta - H^{-1}( dv + \omega ) $, and the Einstein frame metric of the five-dimensional spacetime
   \begin{equation}
\dd s_{\rm BH 5}^{\; 2} = \left(\frac{f^2}{H}\right)^\frac{2}{3} (dv + \omega)^2   -  \left(\frac{H}{f^2}\right)^\frac{1}{3}  \gamma_{mn} \dd x^m \dd x^n \, . 
\end{equation}
For a five-dimensional black hole one takes the asymptotic value of $\frac{f^2}{H}$ to be one. The isometry coordinate $v$ can then be interpreted as the asymptotic time coordinate $t$ of the black hole solution. The typical example is the D1-D5-P BMPV black hole \cite{Breckenridge:1996is} with 
\beq \gamma_{mn} = \delta_{mn}\; , \qquad H = 1 + \frac{Q_3}{|x|^2} \; , \qquad f = \frac{1}{\sqrt{Z_1Z_2}}\; , \quad Z_I = 1 + \frac{Q_I}{|x|^2}\; , \quad \beta = 0 \;, \eeq
and $\omega$ a harmonic 1-form on $\mathds{R}^4$ with anti-selfdual exterior derivative,
\beq \omega = J^+_{mn} \frac{x^m dx^n}{|x|^4} \;, \eeq
 which carries the selfdual angular momentum $J^+_{mn}$.  The microstate geometries associated to such a five-dimensional black hole admit the same asymptotic fall-off for the gauge fields and the metric, and in particular  $\gamma_{mn}$ is asymptotically Euclidean. 

The most important example of a microstate geometry is probably the superstratum solution  \cite{Giusto:2013bda,Bena:2015bea,Bena:2017xbt,Heidmann:2019zws}. This class of solutions is a deformation of a supertube solution with a circular profile \cite{Lunin:2001jy}. The supertube solution is defined for a general closed parametric curve  $f^m(s)$ in $\mathds{R}^4$ such that
\bea \gamma_{mn} &=& \delta_{mn}\; , \qquad H = 1 \; , \qquad f = \frac{1}{\sqrt{Z_1Z_2}}\; ,  \nonumber \\ 
Z_2 &=& 1 + Q_2 \int_0^{2\pi} \frac{ds}{2\pi} \frac{1}{|{ x} - { f}(s)|^2} \; , \qquad 
 Z_1 = 1 + Q_2 \int_0^{2\pi} \frac{ds}{2\pi} \frac{|{ f}^\prime(s)|^2 }{|{ x} - { f}(s)|^2} \;,  \nonumber \\ 
 \omega_m &=& Q_2 \int_0^{2\pi} \frac{ds}{2\pi} \frac{ f_m(s) }{|{ x} - {f}(s)|^2}  \; , \qquad d \beta = ( 1 + \tilde \star) d\omega \ , \eea
with the identification of the coordinates as $v = t$ and $u = y+t$, which defines $t$ as a time coordinate in six dimensions.\footnote{The conventional choice is rather to take $v = \frac{t-y}{2}$, $u = y+t$ and $H=0$ \cite{Giusto:2013bda,Bena:2015bea,Bena:2017xbt,Heidmann:2019zws}, which is equivalent, but we prefer to keep this definition because this change of variables is not a well defined diffeomorphism, as it shifts the time coordinate by a periodic variable.}  The identification of either  $u$ or $y$ as the periodic coordinate along which both the D1 and the D5 branes are wrapped is consistent since a $v$-independent function that is periodic in $u\rightarrow u+2\pi R_y $ is also periodic in $y\rightarrow y+2\pi R_y$. The superstratum solutions are obtained by solving the system iteratively when the functions $Z_I$ are deformed by functions of $u$ and $x^m$, starting from a set of selfdual forms $\Theta^I$ over $\mathds{R}^4$ depending periodically on $u$. The function $H$ then becomes non-trivial and reproduces asymptotically the fall-off of the BMPV black hole solution $H \sim 1 + \frac{Q_3}{|x|^2} + \mathcal{O}(|x|^{-3})$ for some $Q_3$ determined by the original deformation. The superstratum solutions generally inherit from the supertube the property that both five-dimensional angular momenta do not vanish, and that there is a magnetic dipole associated to $\beta$. It has been shown in \cite{Bena:2017xbt} that the selfdual angular momentum can be pushed below the regularity bound for the black hole solution, whereas the anti-selfdual component is non-zero and therefore necessarily remains over the regularity bound since it must vanish for the black hole solution.

For a four-dimensional black hole one must take the metric $\gamma_{mn}$ to be asymptotically Taub-NUT, with an additional compact $S^1$ fibered over $\mathds{R}^3$. The equivalent of the superstratum solution has not been constructed explicitly in this case. 

For a globally hyperbolic metric, one requires moreover that the isometry coordinate $v$ defines a null foliation of spacetime over a Riemannian base space, such that the 1-form field $\omega$ is globally defined over the base space and the pullback metric on a leaf  $\mathcal B$ of the foliation  defines the Riemannian metric 
\beq  \label{Bmetric} \dd s_{\mathcal B}^{\; 2} =  f H \left( du + \beta - \frac{\omega}{ H}\right)^2     + f^{-1} \gamma_{mn} \dd x^m \dd x^n- \frac{f}{H} \omega^2 \,.
\end{equation}
In particular $f H >0$ and $f^{-1} \gamma_{mn} - \frac{f}{H} \omega_m \omega_n >0$ everywhere on $\mathcal B$. Note that this is equivalently a time-like foliation over the same Riemannian base space $\mathcal B$ after the change of variable $u = y+t$.

As we have seen in the preceding section, supersymmetric solutions of the $(2,0)$ theory preserving the same supersymmetry as the BMPV black hole are necessarily solutions in a $(1,0)$ theory. If $D \beta \ne 0$, the hypermultiplet scalar fields must be constant, while if $\beta =0$ they can be arbitrary functions of the coordinate $u$ which do not depend on the four-dimensional base space coordinates $x^m$.  However, one may anticipate that this kind of solution with non-trivial hypermultiplet profile cannot lead to a regular microstate geometry since the hyper-multiplet scalar fields are not constant at asymptotic infinity, but oscillate instead along the circle parametrized by the coordinate $u$. In the absence of hypermultiplets, the system of equations can be solved as in \cite{Giusto:2013rxa,Giusto:2013bda,Bena:2015bea}. Here we shall discuss the case $\beta=0$, and for simplicity we shall assume that $\gamma_{mn}$ does not depend on the coordinate $u$. 
Using the property that one can choose a gauge such that the coefficients $I^{\scalebox{0.6}{$A\hspace{-0.3mm}B$}}_{ij} $ are constants, one concludes directly that the 4-dimensional manifold of metric $\gamma_{mn}$ is hyper-K\"{a}hler.

One can always parametrize the tensor multiplet scalar fields by projective coordinates such that 
\begin{equation}
\cVT^{}_I  = \frac{Z_I}{\sqrt{ (Z,Z)}} \,,\qquad \text{where}\qquad (Z,Z) = \eta^{IJ}_{\scalebox{0.5}{T}} Z_I Z_J \,,
\end{equation}
and therefore
\begin{equation}
M^{\scalebox{0.5}{T}}_{IJ} = \cVT^{}_I \cVT^{}_J  + \delta_{rs}  \mathcal{V}_{\scalebox{0.5}{T}}^r{}_I \mathcal{V}_{\scalebox{0.5}{T}}^s{}_J = 2\frac{Z_I Z_J}{(Z,Z)} - \eta^{{\scalebox{0.5}{T}}}_{IJ} \; , 
\end{equation}
where $\eta^{\scalebox{0.5}{T}}_{IJ}$ is the restriction of the even-selfdual metric to the sublattice of signature $(1,n_{\scalebox{0.5}{T}})$ with $ n_{\scalebox{0.5}{T}}\le n$, introduced in \eqref{etaT}.

Exploiting the remaining freedom in the definition of $Z_I$, one can always define the scaling factor $f$ such that 
\beq f^{2} = \frac{2}{(Z,Z)} \; . \eeq 
Using these definitions, one computes that the coset momentum satisfies 
\beq \delta_{r s} \mathcal{V}_{\scalebox{0.5}{T}}^r{}_I \PT{}_\mu{}^{s} = \partial_\mu \frac{Z_I}{\sqrt{(Z,Z)}} = - \sqrt{2} f^{-1} \delta_{rs}  \mathcal{V}_{\scalebox{0.5}{T}}^{r}{}_I \mathcal{V}_{\scalebox{0.5}{T}}^{s}{}_J \eta_{\scalebox{0.5}{T}}^{JK} \partial_\mu  \frac{Z_K}{(Z,Z)} \; .  \eeq
Using \eqref{eq:susyG} and \eqref{eq:Gr}, one concludes that in an appropriate gauge, the 2-form fields can be written as
\beq B^I = \frac{1}{2\sqrt{2}} \Bigl( 2\eta^{IJ}  \frac{Z_J}{(Z,Z)} du \wedge (dv + \omega) + A^I \wedge du + b^I \Bigr) \; , \eeq
where $A^I$ are 1-forms and $b^I$ 2-forms on the 4-dimensional base space. The field strength $G^I$ are then
\beq G^I =  \frac{1}{2\sqrt{2}} \left[ 2\eta^{IJ} \tilde d \frac{Z_J}{(Z,Z)} \wedge du \wedge (dv + \omega) + \left( \tilde dA^I + \dot{b}^I - 2 \eta^{IJ}  \frac{Z_J}{(Z,Z)} \tilde d\omega \right)  \wedge du + \tilde d {b}^I \right] \; , \eeq
where $ \dot{b}^I = \partial_u b^I$. From the selfduality of $Z_I G^I$ and the anti-selfduality of $(\frac{Z_I Z_J}{(Z,Z)} - \eta_{IJ} ) G^J$, one obtains that
\beq d b^I = \tilde\star d Z_I \; , \qquad (1+ \tilde\star) ( 2 \tilde d\omega -Z_I \Theta^I ) = 0 \;, \qquad \left(\frac{Z_I Z_J}{(Z,Z)} - \eta_{IJ} \right) (1-\tilde\star ) \Theta^J = 0 \; , \eeq
where one defines for convenience 
\beq \Theta^I = \tilde d A^I + \dot{b}^I \; . \eeq 
As in \cite{Giusto:2013rxa}, we further assume that all the $\Theta^I$ are selfdual, such that one gets 
\beq  \tilde d\omega  + \tilde\star \tilde d \omega = Z_I \Theta^I  \;, \qquad  (1-\tilde\star ) \Theta^J = 0 \; . \eeq
Then the last equation that remains is the Einstein equation along the null coordinate $u$
\bea 0 &=& R_{++} - M_{IJ} G^I_{+ij} G^J_+{}^{ij} - 4 \delta_{rs} P_{+}{}^{r} P_{+}{}^{s} - \delta_{rs} \tilde P_+^{AB\,r} \tilde P_{+\,AB}{}^{s} \\
&=& \frac12 \tilde{\star} \, \tilde{d} \, \tilde{\star} \, ( \tilde{d} H + 2 \dot{\omega}) + \frac14 \eta_{IJ} \Theta^I \wedge \Theta^J - (Z , \ddot{Z}) - \tfrac12 ( \dot{Z},\dot{Z})  + \frac{(Z,Z)}{2} \dot{\cV}{}_{\scalebox{0.5}{H}}^{\underline{AB} I}  \dot{\cV}{}_{\scalebox{0.5}{H} \underline{AB} I} \; ,  \nonumber \eea
where ${\cVH}^{\hspace{-1.2mm}\underline{AB}}{}_{I}$ only depends on the $u$ coordinate. 

We therefore retrieve the same system of equation as in  \cite{Giusto:2013bda,Bena:2015bea}  at $\beta=0$, with additional arbitrary functions ${\cVH}^{\hspace{-1.2mm}\underline{AB}}{}_{I}$ that further source the Laplace equation for the function $H$. The system can be solved in steps starting from a given hyper-K\"{a}hler metric $\gamma_{mn}$. One first finds $u$ dependent harmonic functions $Z_I$ on the four-dimensional base. Then, one can solve $d b^I = \tilde\star\,t dZ_I$ for the 2-forms $b^I$ and determines the vector fields $A^I$ such that the $\Theta^I$ are selfdual, up to arbitrary harmonic vectors of selfdual field strength. The 1-form $\omega$ can be solved modulo a harmonic form of anti-selfdual field strength. Finally, one needs to solve the Laplace equation with source for the function $H$. 

However, one can easily convince oneself that there is no regular solution of this kind which is asymptotically $\mathds{R}^{1,4} \times S^1$ or $\mathds{R}^{1,3} \times T^2$. In the asymptotic region $\dot{\omega}$ and $ \eta_{IJ} \Theta^I \wedge \Theta^J $ must fall off rapidly, so that $\tilde \Delta H$ is sourced by a non-zero positive function of the coordinate $u$ which is constant in $x^m$. It follows directly that $H$ is singular and that the solution is not a smooth geometry.

\section{Charge quantization}\label{sec:chargequantization}

The five-dimensional base space metric \eqref{Bmetric} generically does not admit any isometry.  As a Riemannian space, it admits a third homology group $H_3(\mathds{Z})$ of compact cycles, and the flux quantization imposes that for any homology cycle $\Sigma \in H_3(\mathds{Z})$ one has 
\beq \frac{1}{8\sqrt{2}\pi^2} \int_\Sigma G^I = Q^I_\Sigma \in \Lambda_{5,n} \;  ,\eeq
where $\Lambda_{5,n}$ is the even-selfdual lattice of integral vectors of $\SO(5,n)$. For $n=5$ one has  $\Lambda_{5,5}  = I\hspace{-1mm}I_{5,5}$, the standard Lorentzian lattice, while $\Lambda_{5,21}  = I\hspace{-1mm}I_{5,5} \oplus \E_8 \oplus \E_8$ for $n=21$.

For a given choice of primitive cycles $\Sigma_A\in H_3(\mathds{Z})$ such that any $\Sigma = n^A(\Sigma) \Sigma_A $ for some integers $n^A(\Sigma)$, the solution admits the corresponding set of primitive fluxes 
\beq \frac{1}{8\sqrt{2}\pi^2} \int_{\Sigma_A} G^I = Q^I_A \in   \Lambda_{5,n}  \,, \eeq
which must individually be quantized in $\Lambda_{5,n}$. There is always at least one cycle at infinity $\Sigma_\infty$ that defines an $S^3$ (or more generally a Lens space for a four-dimensional black hole) embedded in the four-dimensional base space parametrized by the coordinates  $x^m$. For a five-dimensional black hole solution, this asymptotic $S^3$ is homotopic to the horizon of the black hole, and 
\beq  \frac{1}{8\sqrt{2}\pi^2} \int_{\Sigma_\infty} G^I = Q^I \in  \Lambda_{5,n}  \eeq
are the NS and RR charges of the black hole in $ \Lambda_{5,n} $. The typical example is the D1-D5 system, in which $Q =  (Q_1 {\bf n} , Q_5 {\bf n})$ for a primitive vector ${\bf n}\in \mathds{Z}^5$ that can be chosen to be ${\bf n}=(1,0,0,0,0)$ in the appropriate basis. The so-called large black hole, with a macroscopic horizon area, must also include a momentum $Q_3$ along the circle parametrized by $u$, which can be interpreted as an electric charge for the vector potential $A^3 =  \beta - H^{-1}( dv + \omega ) $ in five dimensions. 

In this paper we consider supersymmetric solutions with a null isometry $\partial_v$.
We will now show that the quantization condition implies that, at a generic point in moduli space, all charges $Q^I_A$ must be proportional to the total charge $Q^I$. 
For this we use the fact that the equations are invariant under $\SO(5,n)$. The stabilizer of the charge $Q$ of a supersymmetric black hole ($Q_1 Q_5>0$) is $\SO(4,n)$. We have seen in the previous section that all solutions with the same supersymmetry as the five-dimensional black hole (respectively four-dimensional) can be obtained from solution of a $(1,0)$ theory with no hypermultiplets. If one starts from a given embedding of the $(1,0)$ theory in the $(2,0)$ or $(2,2)$ theory  in which the scalar fields parametrize $\SO(1,n)$, one can obtain any values of the asymptotic scalar fields using the property that any element of $\cV \in \SO(5,n)$ can be written as $\cV_0 g^{-1} $ with $\cV_0 \in \SO(1,n)$ the tensor multiplet coset representative and  $g \in \SO(4,n)$ a constant group element. All the fields of the theory are then defined from the $(1,0)$ solution and the constant group element $g$. In the notations introduced in Section \ref{EMsection}, one can define this embedding from the projection $\eta^{\scalebox{0.5}{T}}_I{}^J$ such that for a specific choice of metric $\eta_{IJ}^0$ in the $(1,0)$ theory 
\beq \eta^{\scalebox{0.5}{T}}_{IJ} = g_I{}^K g_J{}^L \eta_{KL}^0 \; . \eeq
In particular, the 3-form field strengths are 
\beq G^I = g^I{}_J G^J_0 \; , \eeq
so that the charges associated to the basis of primitive cycles $\Sigma_A$ are 
\beq Q^I_A = \frac{1}{8\sqrt{2}\pi^2} \int_{ \Sigma_A} G^I =  \frac{1}{8\sqrt{2}\pi^2}\int_{ \Sigma_A} g^I{}_J G_0^J =  g^I{}_J Q^J_{A 0} \eeq
where $Q_{A 0}$ is valued in the vector space $  \Lambda_{1,n} \otimes \mathds{R}$. Notice, that supersymmetry indeed requires the solution to be defined in the $(1,0)$ truncation.
The original solution, however, has no physical significance so one should only quantize $Q_A$ and not the original charges $Q_{A 0}$. To understand the set of charges that is allowed, one must find the intersection of 
\beq g( \Lambda_{1,n} \otimes \mathds{R}) \cap \Lambda_{5,n} \ . \eeq
We shall find that for a generic $g$ this intersection is the one-dimensional lattice of charge vectors proportional to $Q$.

 One considers a solution with total charge $ Q = (Q_1 {\bf n} , Q_5 {\bf n})$ for a primitive vector ${\bf n}\in \mathds{Z}^5$ that could be chosen to be ${\bf n}=(1,0,0,0,0)$.  One defines the $\SO(5,5)$ element $g({\bf u}) $ for a real vector ${\bf u}\in \mathds{R}^5$ orthogonal to ${\bf n}$, 
\begin{equation} g({\bf u}) = \left(\begin{array}{cc} \mathds{1} + {\bf u} \times {\bf n}^\intercal & - \frac{Q_1}{Q_5} \bigl( {\bf u} \times {\bf n}^\intercal - ({\bf n} + |{\bf n}|^2 {\bf u}) \times {\bf u}^\intercal \bigr) \\ 0 &  \mathds{1}  - {\bf n} \times  {\bf u}^\intercal \end{array} \right) \ ,
\end{equation}
such that it stabilizes the charge $Q$,
\begin{equation}   g({\bf u})  \left( \begin{array}{c} Q_1 {\bf n} \\ Q_5 {\bf n}\end{array}\right)  = \left( \begin{array}{c} Q_1 {\bf n} \\ Q_5 {\bf n}\end{array}\right) \ . \end{equation} 
The charges $Q_{A 0}$ of the $(1,0)$ truncation belong to the real extension of the lattice $\Lambda_{1,5}  =  I\hspace{-1mm}I_{1,1} \oplus A_1^{\; 4}$, where the first factor is parametrized by  $(q_1 {\bf n} , q_5 {\bf n})  \in   I\hspace{-1mm}I_{1,1} $, while one can parametrize the $A_1^{\; 4}$ component by a vector $({\bf q},-{\bf q})$ with ${\bf q} \in \mathds{Z}^4$ orthogonal to ${\bf n}$. We shall consider $q_1$, $q_5$ and ${\bf q}$ to be real since they do not define the quantized physical charges. The physical charge is defined after the action of $g({\bf u})$ as
\begin{equation}   g({\bf u})  \left( \begin{array}{c} q_1 {\bf n} + {\bf q} \\ q_5 {\bf n}- {\bf q} \end{array}\right)  = \left( \begin{array}{c} (q_1 +\tfrac{Q_1}{Q_5} {\bf u} \cdot {\bf q}) {\bf n} + {\bf q} + {\bf u}  |{\bf n}|^2 \bigl( q_1 -\tfrac{Q_1}{Q_5} (q_5- {\bf u} \cdot {\bf q} ) \bigr) \\  (q_5+ {\bf u} \cdot {\bf q} ) {\bf n} - {\bf q} \end{array}\right) \ . \end{equation} 
For the resulting charge to be in the lattice $\Lambda_{5,5}$, one needs that 
\begin{equation}  {\bf q} \in \mathds{Z}^5 \; {\rm mod} \; {\bf n}\;, \quad  q_5+{\bf u} \cdot {\bf q} =q_5^\prime \in \mathds{Z} \; , \quad q_1 +\tfrac{Q_1}{Q_5} {\bf u} \cdot {\bf q}=q_1^\prime \in \mathds{Z} \ ,    \end{equation} 
and therefore also
\begin{equation} 
 {\bf u}  |{\bf n}|^2 \left( q^\prime_1 -\tfrac{Q_1}{Q_5} q^\prime_5 +\tfrac{Q_1}{Q_5}{\bf u} \cdot {\bf q}  \right) \in \mathds{Z}^5  \ .  \end{equation} 
If ${\bf u}$ is generic in $\mathds{R}^4$, there is no vector ${\bf x} \in \mathds{Q}^4$ such that ${\bf x} \cdot {\bf u} \in \mathds{Q}$, except ${\bf x} = 0$. Thus, if one component of $ {\bf u} \bigl( q^\prime_1 -\tfrac{Q_1}{Q_5} q^\prime_5 +\tfrac{Q_1}{Q_5}{\bf u} \cdot {\bf q} \bigr)$ is an integer, then the others cannot be integer unless they all vanish. One finds therefore that the only solution is the trivial one for which ${\bf q} = 0$ and $Q_5 q_1 = Q_1 q_5$.
This means that the lattice of allowed charges is generated by $Q / {\rm gcd}(Q_1,Q_5)$. 

One can similarly find an $\SO(5,21)$ group element $g({\bf u},{\bf v})$ stabilizing $Q$ with ${\bf u} \in \mathds{R}^5$ such that ${\bf u} \cdot {\bf n}=0$ and ${\bf v} \in (\E_8\oplus \E_8) \otimes \mathds{R}$, so that 
\bea  g^\prime({\bf u},{\bf v}) \left( \begin{array}{c}  q_1 {\bf n} + {\bf q} \\ q_5 {\bf n}- {\bf q} \\ {\bf p} \end{array}\right)  &=& \left( \begin{array}{c} q_1 {\bf n} + {\bf q} + {\bf u} \bigl( k({\bf v} ,{\bf p}) - \tfrac12 k^{}({\bf v},{\bf v}) {\bf u} \cdot {\bf q} \bigr) \\
q_5 {\bf n} - {\bf q} \\
{\bf p} - {\bf v} \, ({\bf u} \cdot {\bf q})\end{array}\right) \nonumber\\
&=& \left( \begin{array}{c} q_1 {\bf n} + {\bf q} + {\bf u} \bigl( k({\bf v} , {\bf p}^\prime) + \tfrac12 k^{}({\bf v},{\bf v}) {\bf u} \cdot {\bf q} \bigr) \\
q_5 {\bf n} - {\bf q} \\
{\bf p}^\prime \end{array}\right) \ . \eea
Here $k$ denotes the Killing Cartan form on the root lattice $\E_8 \oplus \E_8$. 
One obtains the similar condition that 
\beq {\bf u} \left( k({\bf v} ,{\bf p}^\prime) + \tfrac12 k^{}({\bf v},{\bf v}) {\bf u} \cdot {\bf q} \right) \in \mathds{Z}^5 \; , \eeq 
with ${\bf p}^\prime \in \E_8\oplus \E_8$ and ${\bf q} \in \mathds{Z}^4$, so that for generic ${\bf u}$ and ${\bf v}$ the only allowed solution is ${\bf q} = {\bf p}=0$. Combining two elements of the form $g({\bf u}) g^\prime({\bf u}^\prime,{\bf v})$ as defined above, one arrives at the conclusion that the only allowed charges are proportional to the total charge $Q / {\rm gcd}(Q_1,Q_5)$.

\section{Conclusions}
In this paper we studied supersymmetric solutions of $(2,0)$ supergravity in six dimensions. The spinors are defined in a real $\mathds{R}^{4\times 4}$ vector space inside the  $\mathds{C}^{4\times 4}$ tensor product representation of $SU^*(4)$ and $\USp(4)$. A Killing spinor is at least rank 2 as such a four by four matrix, and for any given rank 2 spinor one can decompose $\mathds{R}^{4\times 4}$ into four $\mathds{R}^{2\times 2}$ orthogonal components using \eqref{eq:spinorprojectionA} and \eqref{eq:spinorprojectionB}. Supersymmetric black hole solutions in $\mathds{R}^{1,4} \times S^1$ and  $\mathds{R}^{1,3} \times T^2$ with four Killing spinors have their four Killing spinors in the same $\mathds{R}^{2\times 2}$ subspace. We proved that all supersymmetric solutions  with four Killing spinors of this type are also solutions of a $(1,0)$ theory with the same preserved supersymmetries.

This result has a direct consequence for the search of microstate geometries within Mathur's proposal. We exhibit that smooth solutions with the same four supersymmetries as BPS black holes are also solutions  of a $(1,0)$ theory with hypermultiplet scalar fields that only depend on the coordinate parametrising the circle at infinity. It follows therefore from the asymptotic behaviour of these solutions that smooth microstate geometries only exist if the hypermultiplet scalar fields are constant. We conclude that all such solutions can be obtained from solutions of standard $(1,0)$ supergravity involving tensor multiplets only. 

The same proof applies to maximal supergravity if one disregards the vector fields. In principle it is possible that solutions involving the vector fields would allow for more general solutions than the ones that can be constructed in a $(1,0)$ theory. But one does not expect the black hole microstate realisation of supersymmetric D1-D5-P black holes to be qualitatively different on $T^4$ and on K3, so it is difficult to believe that including the vector fields of $(2,2)$ supergravity could drastically change the conclusion.

Moreover, we infer from this result that microstate geometries with appropriately quantised fluxes in six dimensions cannot exist at a generic point of the $(2,0)$ supergravity moduli space unless all the three-form fluxes are proportional to the total flux of the solution. Multi-cycle solutions that define bound states, in the sense that one cannot move the various cycles apart without changing the moduli or the flux they carry, only exist when their fluxes are linearly independent \cite{Bena:2007kg,Giusto:2004id,Bena:2005va,Berglund:2005vb,Bena:2006kb,Bena:2007qc,Bianchi:2017bxl,Heidmann:2017cxt,Bena:2017fvm}. 
Supersymmetric solutions where the various cycles can freely be moved in spacetime are understood to describe multi-centre black holes of the Papapetrou--Majumdar type \cite{Papaetrou:1947ib,Majumdar:1947eu}, and are therefore not relevant for microstate geometries of a single black hole solution.
Furthermore, single-centre black holes exist everywhere in moduli space \cite{Ferrara:1996um}, the same must hence be true for their microstate geometries.
 We conclude therefore that all microstate geometries which are relevant for the description of single-centre supersymmetric black holes can only carry one single cycle. 

This therefore rules out the possibility that multi-bubble solutions are admissible supersymmetric black hole microstate geometries.
Instead, it suggests that one must concentrate on supertube-like solutions as the superstratum. Our findings are also consistent with previous results on the D1-D5 orbifold conformal field theory, since
only supertube-like solutions seem to admit a holographic description within this framework \cite{Galliani:2017jlg,Tormo:2018fnt,Giusto:2018ovt,Giusto:2019qig}.

\subsection*{Acknowledgements}

We would like to thank Iosif Bena,  Massimo Bianchi, Jose F.\ Morales, Andrea Puhm,  Ashoke Sen and Nick Warner for useful discussions. This work was partially supported by the ANR grant Black-dS-String (ANR-16-CE31-0004).
The work of S.L.\ is also supported by the ERC Starting Grant 679278 Emergent-BH.


\appendix
\noindent

\section{Conventions}\label{app:conventions}
The \(\USp(4)\) invariant tensor \(\omega_{AB}\) satisfies
\begin{equation}
\omega_{AC}\omega^{BC} = \delta_A^B \,.
\end{equation}
It is used to raise and lower indices according to
\begin{equation} \label{ConvRL} 
X^A = \omega^{AB} X_B \,,\qquad X_A = X^B \omega_{BA} \,.
\end{equation}
\(\USp(4)\) is locally isomorphic to \(\SO(5)\).
In particular, its antisymmetric traceless representation agrees with the vector representation of \(\SO(5)\).
Explicitly, every vector \(A^a\), \(a = 1, \dots 5\), of \(\SO(5)\) can be transformed into an antisymmetric symplectic traceless \(\USp(4)\) tensor \(A^{AB}\) with the help of the \(\Gamma\)-matrices of \(\SO(5)\), \ie 
\begin{equation}\label{eq:appso5gamma}
{A_A}^B = \frac12 A^a {\left(\Gamma_a\right)_A}^B \,.
\end{equation}
Therefore, as a direct consequence of the anticommutation relation of the \(\Gamma^a\), any two antisymmetric tensors \(A^{AB}\) and \(B^{AB}\) of \(\USp(4)\) satisfy the identity
\begin{equation}\label{eq:appusp4clifford}
A_{AC} B^{BC} + B_{AC} A^{BC} = \tfrac12 A_{CD} B^{CD} \delta_A^B \,.
\end{equation}

We use a `mostly negative' spacetime signature,
\begin{equation}
\eta_{\alpha\beta} = \mathrm{diag}(+1, -1,-1,-1,-1,-1) \,.
\end{equation}
The chirality projector $\gamma_7$ is given by
\begin{equation}
\gamma_7 = \gamma_{012345} \,,
\end{equation}
and the supersymmetry parameter \(\epsilon^A\) is anti-chiral,
\begin{equation}
\gamma_7 \epsilon^A = - \epsilon^A \,.
\end{equation}
Hodge duality acts on the \(\gamma\)-matrices as
\begin{equation}\label{eq:dualgamma}
\gamma^{\mu_1 \dots \mu_n} = \frac{(-1)^{\lfloor \frac{n}{2} \rfloor }}{(6-n)!} \varepsilon^{\mu_1 \dots \mu_n \nu_1 \dots \nu_p} \gamma_{\nu_1 \dots \nu_p} \gamma_7 \,.
\end{equation}
Moreover, for a selfdual three-form $G^+ = \star G^+$ it follows that
\begin{equation}\label{eq:6dselfdual}
G^+_{\mu\nu\rho} \gamma^{\mu\nu\rho} \epsilon^A = 0 \,.
\end{equation}

The general expansion of an (anti-)selfdual three-form \(G^\pm = \pm \star G^\pm\) with respect to the null frame \eqref{eq:nullbasis} reads 
\begin{equation}\label{eq:Gdecomp}
G^\pm = e^+ \wedge A^\mp + e^- \wedge B^\pm + (1\pm \star) e^+ \wedge e^- \wedge C \,,
\end{equation}
where \(A^\pm = \frac12 A^\pm_{ij} e^i \wedge e^j\) and \(B^\pm = \frac12 B^\pm_{ij} e^i \wedge e^j\) are (anti-)selfdual two-forms with respect to \(-\delta_{ij}\), i.e.~\(\star_4 A^\pm = \pm A^\pm\) and \(C = C_i e^i\).

In the null frame \eqref{eq:nullbasis} we introduce the four-dimensional chirality matrix:
\begin{equation}
\gamma_\ast = \gamma_{1234} \,.
\end{equation}
A relation analogous to \eqref{eq:dualgamma} holds for the four-dimensional \(\gamma\)-matrices.
Moreover, the projection \eqref{eq:spinorprojectionB} is equivalent to
\begin{equation}\label{eq:app4dchiral}
\gamma_\ast \epsilon^A = \epsilon^A \,.
\end{equation}
Thus it follows that
\begin{equation}\label{eq:4dselfdual}
A^+_{ij} \gamma^{ij} \epsilon^A = 0 \,,
\end{equation}
 for every selfdual two-form \(A^+\).

\section{Integrability conditions}\label{app:integrability}

The gravitini Killing spinor equation \eqref{eq:psivariation} implies the integrability condition
\begin{equation}\label{eq:killingintegrability}
\Bigl(   R_{\mu\nu\kappa\lambda} \delta^A_B \gamma^{\kappa\lambda} + 8 P_{[\mu \, BC\, r} P^{AC\, r}_{\nu]} + 4 D_{[\mu} G^{AC}_{\nu]\kappa\lambda} \omega_{CB} \gamma^{\kappa\lambda}  - 2 G^{AC}_{\kappa\lambda[\mu} G^{DE}_{\nu]\gamma\delta} \omega_{CD} \omega_{EB} \gamma^{\kappa\lambda} \gamma^{\gamma\delta}  \Bigr )  \epsilon^B  = 0 \,.
\end{equation}
After contracting this with $\gamma^\mu$ we obtain
\begin{equation}\begin{aligned}
&\left[2 \omega^{AB} E_{\mu\nu} \gamma^\nu + \frac23 \cV^{AB}{}_I E^I_{\mu\nu\kappa\lambda}\gamma^{\nu\kappa\lambda} + 4 \cV^{AB}{}_I (\star E^I)_{\mu\nu} \gamma^\nu \right] \omega_{BC}  \epsilon^C \\
&\quad+ i \left[8 P^{AC\,r }_\mu \omega_{CB} + \frac13G^r_{\nu\kappa\lambda} \gamma^{\nu\kappa\lambda} \gamma_\mu \delta^A_B \right] \delta \chi^{B\,r} = 0 \,,
\end{aligned}\end{equation}
where $E_{\mu\nu}$ and $E^I$ are defined in \eqref{eq:einstein} and \eqref{eq:3formeom} and represent the Einstein equations and the equations of motion of the two-form fields.
Therefore, if the  Bianchi identity \eqref{eq:bianchi} for $G^{AB}$ is satisfied, the integrability condition
reduces to
\begin{equation}
E_{\mu\nu} \gamma^\nu \epsilon^A = 0 \,,
\end{equation}
and by multiplying with $\bar \epsilon^B \gamma^{\kappa\lambda}$ from the left we find
\begin{equation}
E_{\mu [\kappa} V_{\lambda]} = 0 \,,
\end{equation}
or equivalently
\begin{equation}
E_{\mu-} = E_{\mu i} = 0 \,.
\end{equation}
So all but the $++$ component of the Einstein equations follow from the Killing spinor equation.

There is another integrability condition for the tensorini variation \eqref{eq:deltachi} which reads
\begin{equation}
-i \gamma^\mu D_\mu \delta \chi^{A\,r} = \left[E^{AB\,r} \omega_{BC} +  \frac12 \cV^r{}_I (\star E^I)_{\mu\nu} \gamma^{\mu\nu} \delta^A_C \right] \epsilon^C + \frac{i}{6} G^{AB}_{\mu\nu\rho} \omega_{BC} \gamma^{\mu\nu\rho} \delta\chi^{C\,r} = 0 \,,
\end{equation}
with $E^{AB\,r}$ and $E^I$ from \eqref{eq:scalareom} and \eqref{eq:3formeom}.
Consequently, if the Bianchi identity \eqref{eq:bianchi} for $G^r$ holds, also the scalar equation of motion is implied by the Killing spinor equations.



\clearpage
\bibliography{references}  

\bibliographystyle{utphys}

\end{document}